# Motion Hologram: Jointly optimized hologram generation and motion planning for photorealistic and speckle-free 3D displays via reinforcement learning

Zhenxing Dong[1], Yuye Ling[1,*], Yan Li[1], and Yikai Su[2]

[1] Department of Electronic Engineering, Shanghai Jiao Tong University, Shanghai 200240, China
[2] State Key Lab of Advanced Optical Communication Systems and Networks, Shanghai Jiao Tong University, Shanghai 200240, China
[*] Corresponding authors: yuye.ling@sjtu.edu.cn

## Abstract

Holography is capable of rendering three-dimensional scenes with full-depth control, and delivering transformative experiences across numerous domains, including virtual and augmented reality, education, and communication. However, traditional holography presents 3D scenes with unnatural defocus and severe speckles due to the limited space bandwidth product of the spatial light modulator (SLM). Here, we introduce Motion Hologram, a novel holographic technique to accurately portray photorealistic and speckle-free 3D scenes, by leveraging a single hologram and learnable motion trajectory, which are jointly optimized within the deep reinforcement learning framework. Specifically, we experimentally demonstrated the proposed technique could achieve a 4~5 dB PSNR improvement of focal stacks in comparison with traditional holography and could successfully depict speckle-free, high-fidelity, and full-color 3D displays using only a commercial SLM for the first time. We believe the proposed method promises a new form of holographic displays that will offer immersive viewing experiences for audiences.

# Introduction

Since its first inception in 1947 by Dennis Gabor, holography (*1*) has been hailed as the "Holy Grail" of information display. Unlike existing three-dimensional (3D) display techniques such as stereoscopic displays (*2*), volumetric displays (*3*), and light-field displays (*4*), holographic displays (*5*) offer pixel-level depth control, aberration correction, and a compact form factor, which promises an immersive and photorealistic viewing experience to the audience. Computer-generated holography (CGH) (*6, 7*) is one of the most popular and prospective technique in holographic displays. It digitally synthesizes holograms using computational methods and dynamically reproduces the wavefronts of 3D virtual objects by coherently illuminating the spatial light modulators (SLMs).

Recently, CGH algorithms (*8-14*) have achieved remarkable breakthroughs with the aid of exploding artificial intelligence (AI) technology, especially for its application in augmented reality (AR) and virtual reality (VR). Despite these advancements, current CGH still faces significant challenges that restrict its usage in practice owing to hardware limitations. Firstly, a single 2D SLM can hardly be used to accurately render high-fidelity 3D scenes with natural focus cue in traditional holography due to limited bandwidth (*11*). The 3D scenes reconstructed by traditional CGH systems often exhibit unnatural artifacts and laser speckles, which significantly compromise the viewer's experience. Secondly, the current optical system design lacks a unified paradigm that can effectively integrate it with hologram generation algorithms to further unlock the full potential of CGH to deliver photorealistic 3D images.

To address the first issue of limited bandwidth, several techniques (*15-23*) have been proposed and most of them are rooted in the concept of multiplexing. First of all, one straightforward implementation is to spatially tile multiple SLMs (*15, 24, 25*) to enhance the space bandwidth product, thereby achieving realistic 3D displays. Nonetheless, this approach would inevitably increase system complexity and computational overheads. Secondly, time-multiplexed holography (*21, 22*), which displays multiple frames with distinct speckle patterns in rapid succession to break the temporal coherence of light source, allows the human viewers to perceive realistic 3D displays without artifacts through subjective visual averaging. However, most commercially available SLMs possess a refresh rate of merely 60 Hz, which is inadequate to meet the requirements of time-multiplexed methods when being operated at video rate. While some researchers (*16, 17, 23*) have proposed to utilize high-refresh rate devices such as ferro-electronic liquid crystal on silicon (FLCOS) and micro-electromechanical systems (MEMS), these systems are less mature in performance due to the limited pixel counts and shallow bit depths (*26*). Additionally, Nam et al.(*19*) introduced a polarization-multiplexed metasurface to enhance the space bandwidth product of SLM, despite the final displayed multi-plane images only exhibit marginal quality improvement.

Secondly, most hardware systems and software algorithms are designed separately in traditional holography, or more generally in optical tasks, which fails to fully unlock the potential of the hardware (*27-29*). A novel and promising approach is to solve the systems and algorithms within an end-to-end differentiable framework in the computational display and imaging fields (*30-34*). For instance, Tseng et al. (*33*) devised a differentiable holographic image formation model that jointly learned a diffractive optical element (DOE) with the hologram pattern to expand the field-of-view by 64 times. Despite these inspiring attempts, this differentiable architecture might not be the ideal and universal solution. Firstly, it requires the hardware configurations to be differentiable; however, not all system components can be described by differentiable mathematical models, such as scanning devices and illumination sources (*35-37*). Secondly, the current pipeline is mostly demonstrated in single hardware optimization tasks, such as the design of DOE and metasurface, despite of the fact that real-world systems are more complicated and usually involves multiple devices. (*38*). Recently, reinforcement learning (RL) (*39*) has emerged as an effective way to optimize the complex and non-differentiable scenarios (*40-42*). Specifically, RL is a reward-based learning method (*43*) that identifies the optimal strategy in the parameter space through trial and error without the need of gradient information (*44*), which appears as a promising candidate to co-design a complicated system.

Here, we proposed a novel CGH technique called Motion Hologram to disrupt the spatial coherence of light source using reinforcement learning, which can thus render photorealistic and speckle-free 3D scenes by globally designing the holographic system. We reappraised the role of system motion, which was commonly believed detrimental in imaging and display, and conceptualized that the inherent speckles in holography could be neutralized and photorealistic 3D display could be achieved by spatially multiplexing the holograms via proper system motions. Specifically, a novel system design paradigm that leverages reinforcement learning to jointly generate the phase-only hologram and system's motion trajectory were introduced, which promises an unparalleled viewing experience. To showcase the superior performance of the proposed technique, we conducted numerical simulations and experimental reconstructions alongside traditional and time-multiplexed holography methods. We developed a benchtop prototype and demonstrated that a 4~5 dB improvement in the PSNR is achieved compared with traditional holography, which also surpasses that of state-of-the-art time-multiplexed holography. Furthermore, our method is capable of accurately reconstructing full-color and high-fidelity multi-plane images with realistic defocus in visual effects. We believe the proposed method promises a new form of holographic displays that reconstructs photorealistic and speckle-free 3D scenes and provides a novel blueprint for computational design in various research areas.

# Results

## Motion Hologram

Traditional holographic display employs a collimated laser beam to illuminate a phase-only SLM, and thus an optical field is generated in the SLM plane. This field is then modulated by the hologram and propagated over a specific distance to reconstruct desired 2D images or 3D image stacks. It is common to utilize the angular spectrum method (ASM) (*45, 46*) to calculate the propagation of complex waves from the SLM plane to the target plane, whose expression is given as follows:

$$u_{\text{recon}}(x, y) = \iint \mathcal{F}(e^{i\phi(x,y,\lambda)}) \mathcal{H}(f_x, f_y, \lambda, Z) e^{i2\pi(f_x x + f_y y)} df_x df_y, \quad (1)$$

$$\mathcal{H}(f_x, f_y, \lambda, Z) = \begin{cases} e^{i\frac{2\pi}{\lambda}\sqrt{1-(\lambda f_x)^2-(\lambda f_y)^2}z}, & \text{if } \sqrt{f_x^2 + f_y^2} < \frac{1}{\lambda}, \\ 0 & \text{otherwise} \end{cases} \quad (2)$$

where $u_{\text{recon}}(x, y)$ is the optical field in the target plane, $\phi(x, y, \lambda)$ is the phase-only hologram loaded on the SLM. $\lambda$ is the wavelength, $f_x, f_y$ are spatial frequencies, $Z$ is the propagation distance between SLM and target plane, $\mathcal{H}\{\cdot\}$ is ASM's transfer function and $\mathcal{F}\{\cdot\}$ denotes the Fourier transform. The reconstructed images $|u_{\text{recon}}(x, y)|$ can be obtained by taking the absolute value of $u_{\text{recon}}(x, y)$.

For 3D displays, ASM can also be used to solve the inverse problem of optimizing $\phi$ for a focal stack intensity $I_{\text{target}}^j$ located at the set of distances $z^j, j = 1, 2, \ldots, J$. The objective is to solve the following optimization problem:

$$\phi = \underset{\phi}{\arg\min} \mathcal{L}\left(|u_{\text{recon}}^j|, \sqrt{I_{\text{target}}^j}\right), \quad (3)$$

where $I_{\text{target}}^j$ is usually rendered based on incoherent wave propagation (*16, 21*). In traditional holography, a single hologram is used to directly control the 3D volume of light with natural focus cues. However, this task often requires a larger space bandwidth product than what the SLM can offer, which leads to 3D reconstructions with uncontrollable speckle noise and unnatural ringing artifacts.

To address this challenge, our strategy is to generate multiple replicas of the image with identical spatial speckle patterns, such that the speckles could be reduced when these images are spatially overlaid and averaged. Differentiating from the aforementioned space- or time-multiplexed holography which requires multiple or high-speed SLMs to generate different patterns at the same location to smooth out the speckles, we aim to utilize motion to synthesize a specific form of "motion speckle": by superimposing and averaging each replica along the motion trajectory, we could effectively eliminate spatial speckles and display high-quality 3D scenes with realistic defocus. Furthermore, unlike previous works on CGH algorithms, we develop an innovative design paradigm that leverages deep reinforcement learning to jointly optimize the hologram generation and motion planning to deliver unparalleled display quality in an end-to-end manner.

The overall framework is shown in Fig. 1(a). The deep agent $\pi_\theta$ interacts with the environment, in this case the Motion Holography, in a sequence of actions $a$, states $s$, and rewards $r$. Specifically, the deep agent $\pi_\theta$ consists of two networks: an actor network and a critic network. We define the action $a$ as the combination of a specific direction ($d$) and a number of pixels ($p$) of each movement performed by SLM and the state $s$ represents the current position of the SLM and the number of steps ($k$).

In the joint optimization stage, we first initialize the state $s_{k-1}$ and feed it to both the actor and critic networks. The actor network is expected to generate the next action $a_k$ for Motion Holography based on the received state $s_{k-1}$. Next, the motion planning of the SLM is updated according to the newly generated action $a_k$, and the phase-only hologram $\phi_{\text{motion}}$ is re-optimized based on the latest motion trajectory in Motion Holography. Once the motion hologram $\phi_{\text{motion}}$ is optimized, the immediate reward $r_k$ and the new state $s_k$ are fed to the critic network and the actor network to collect the next round of action, reward, and state. After iteration between the deep agent $\pi_\theta$ and Motion Holography, a complete motion trajectory is thus obtained. The motion trajectory and the corresponding cumulative reward $R$ can then be expressed as follows:

$$Trajectory = \{a_1, a_2, a_3, \ldots, a_K\}, \quad (4)$$

$$R = r_1 + \gamma r_2 + \gamma^2 r_3 + \cdots + \gamma^{K-1} r_K = \sum_{k=1}^{K} \gamma^{k-1} r_k, \quad (5)$$

where $K$ is the total step of motion termination and $\gamma \in [0, 1]$ is a discount factor, which aims to reduce the impact of future rewards. The goal of agent is to strategically choose actions to maximize the expected cumulative reward $R$. Therefore, we solve the following optimization problem in the proposed framework:

$$\phi_{\text{motion}}, Trajectory = \underset{\phi_{\text{motion}}, Trajectory}{\operatorname{argmax}} \sum_{k=1}^{K} \gamma^{k-1} r_k. \quad (6)$$

The detailed scheme of Motion Holography is presented in Fig. 1(b). Firstly, the hologram is digitally shifted to a specific position according to the motion trajectory, which emulates the physical movement of the SLM in the real world, to create the spatial replicas of the hologram. Then, each hologram is propagated in free space using ASM model or camera-in-the-loop (CITL) model to reconstruct a 3D scene with the same motion speckle pattern but at slightly shifted spatial locations. Finally, the reconstruction result $|u_{\text{motion}}^j|$ is obtained by superimposing and averaging all replicas. We optimize the motion hologram using stochastic gradient descent (SGD) solver to solve the following optimization problem:

$$|u_{\text{motion}}^j| = \frac{1}{K} \sum_k |f_{\text{model}}\{\text{Trajectory}(\phi_{\text{motion}}), z^j\}|, \quad (7)$$

$$\phi_{\text{motion}} = \underset{\phi}{\operatorname{argmin}} \mathcal{L}\left(|u_{\text{motion}}^j|, \sqrt{I_{\text{target}}^j}\right), \quad (8)$$

where we utilize the DeepFocus (*47*) network to retrieve multi-plane images stack $I_{\text{target}}^j$ with natural focus cues from RGB-depth (RGBD) images. More detailed information can be found in the "Methods" section and Supplementary Materials.

The details of deep agent $\pi_\theta$ architectures are showcased in Fig. 1(c), which are implemented using multi-layer perceptrons (MLPs). The actor network learns a policy to make decisions, while the critic network outputs the value $v_k$ to evaluate the actions employed by the actor based on the reward. The immediate reward $r_k$ as shown in Fig. 1(d), can be evaluated by:

$$r_k = r_k^{\text{all}} + W \times r_k^{\text{in-focus}}, \quad (9)$$

where $r_k^{\text{all}}$ represents the improvement in PSNR of the entire reconstructed 3D scene after each action is taken, $r_k^{\text{in-focus}}$ is a metric to measure the increment in PSNR after each action only in the focal region considering that the reconstruction quality in this area remains our primary target, and $W$ gives the weight assigned to $r_k^{\text{in-focus}}$.

Here, the proximal policy optimization (PPO) (*48*) method is used to maximize the expected reward $R$. Specifically, the framework is jointly optimized in an automated differentiation software (*49*), and then backpropagated through the neural network to update $\pi_\theta$. The network architecture and optimization process are detailed in "Methods" section.

## Simulation results

We first compared the focal stacks obtained by traditional holography, time-multiplexed holography, and proposed holography in simulation, as shown in Fig. 2. It is observed that the 3D scenes reconstructed by traditional holography present significant speckles and chromatic aberrations, no matter where the camera is focused (Fig. 2, A and D). In addition, a relatively lower-resolution rendering for in-focus regions and unnatural ringing patterns in out-of-focus regions (the zoom-in images in Fig. 2, A and D) lead to an unpleasant viewing experience for the audience. In contrast, the proposed method leverages motion to synthesize unique "motion speckles" (Fig. S1) to reconstruct high-quality and speckle-free 3D scenes with higher-resolution and photorealistic contents (Fig.2, B and E) via temporal averaging. It is worth noting that the proposed holography can deliver visual effects comparable to those of the current state-of-the-art time-multiplexed holography (Fig.2 C and F).

The quantitative results are provided in Fig. 3(a) for comparison purpose. The peak signal-to-noise ratio (PSNR) of images reconstructed using traditional holography is 23.22 dB, while the proposed holography achieves PSNR of 33.22 dB, which represents a significant improvement of 10 dB. Similarly, the structural similarity index measure (SSIM) of focal stacks using the proposed holography is 0.92, which outperforms the traditional holography by a large margin of 0.47 SSIM. We further conducted and compared two-frame and four-frame time-multiplexed holography to verify the superiority of the proposed method. The results reveal that our method significantly outperforms the two-frame time-multiplexed holography in PSNR and SSIM by 5.36 dB and 0.17, respectively. Additionally, the proposed holography also demonstrates display quality that is equivalent to that of the four-frame time-multiplexed holography in the quantitative metrics.

## Model analysis

In this section, we conducted a detailed model analysis to examine the role of each action in the optimized trajectory and demonstrate the effectiveness of the reward function. Firstly, the cumulative rewards obtained at each step along the optimal trajectory are shown in Fig. 3(b). It is clear that the cumulative rewards generally increase with the number of steps; however, the correlation is not perfect and the benefits tend to be marginal under certain circumstances. For example, after taking the third and sixth steps, the total reward is calculated to be 10.77 dB and 17.09 dB, which equates to a reduction of 0.38 dB and 0.07 dB from the cumulative rewards of the former steps, respectively. This interesting phenomenon occurs because the optimization goal of the proposed joint framework is to maximize the cumulative reward after taking the terminal step rather than the incremental gain at each individual step, as illustrated in Equation 6.

Secondly, we examined the impact of $r_k^{\text{in-focus}}$ on the final trajectory. Fig. 3(c) illustrates the trajectories obtained with and without $r_k^{\text{in-focus}}$, which are labelled as optimized trajectory and non-optimized trajectory, respectively. It is worth noting that the non-optimized trajectory for SLM is only extended in one direction, which might cause motion blur and lower sharpness in the reconstructed 3D scene. To further analyze this result, we conducted quantitative and qualitative analyses, which are presented in Fig. 3(d) and Fig. 3(e). When $r_n^{\text{in-focus}}$ is not enforced as an additional reward function, the total reward obtained by using non-optimized trajectory is 10.76 dB, which is even 1.13 dB higher. However, a simple visual inspection would tell a difference story: despite of its relatively higher quantitative metrics, the 3D reconstructed scene based on non-optimized trajectory not only exhibits undesirable motion blur in the in-focus regions (as indicated by blue arrows) but also introduces ringing artifacts at the boundaries between the in-focus and out-of-focus regions (as indicated by red arrows). In Fig. 3 (d), when $r_k^{\text{in-focus}}$ is incorporated, the proposed method achieves a higher total reward with optimized trajectory than with non-optimized trajectory. Furthermore, the proposed method successfully renders high-fidelity and photorealistic 3D scenes without motion blur and ringing artifacts using optimized trajectory in Fig. 3(e).

**Experimental results**

We conducted real-world optical experiments to further validate the effectiveness of the proposed method. A benchtop holographic display was built and calibrated for experimental verification (Fig. 6 for a setup visualization and refer to Methods for system calibration details). Fig 4 and S13-S15 present experimentally captured images of multi-plane full-color scenes, each is focused either at a near or a far distance. Additionally, we provide the videos of static and dynamic scenes to further demonstrate the superior performance of the proposed method in the Supplementary Video.

We qualitatively and quantitatively assessed and compared the experimental results with those of traditional holography, time-multiplexed holography, and proposed holography. In line with the simulation results, the proposed holography not only perfectly generates speckle-free and higher-contrast multi-plane image stacks (Fig.4 B and E), but also naturally portrays 3D defocusing effects over the entire volume (the Supplementary Video), which ensures a more realistic viewing experience. In comparison, traditional holography suffers from speckle noise and low contrast in the reconstructed 3D scenes (Fig. 4 A and D). Moreover, in certain scenes, such as the cave in Fig. 4 D and the flowers in Fig. S14, traditional holography fails to accurately depict 3D defocus behaviour, which causes vergence-accommodation conflict for audiences. Quantitatively, the proposed holography achieves PSNR of 25.62 dB and 24.17 dB in the reconstructed Bamboo and Bunny scenes, which showcase robust and significant improvements of 3.92 dB and 4.8 dB over traditional holography, respectively. To our best knowledge, the proposed holography represents the highest performance reported thus far in 3D holography using a single SLM and a single laser source. Furthermore, the proposed method also meaningfully matches the performance of time-multiplexed holography in both visual quality and evaluation metrics.

# Discussion

In summary, we introduced a novel CGH technique, Motion Hologram, to achieve photorealistic and speckle-free 3D displays. This innovative form of holography leverages motion to disrupt the spatial coherence of the light source and synthesize unique speckle patterns, which will then be averaged over all reconstructed images along the motion trajectory to attain this purpose. A reinforcement learning-based framework was proposed to jointly optimize the hologram generation and motion planning for the first time in an end-to-end manner to ensure the excellent display quality of the contents. Extensive simulation and experimental results in addition to a thorough analysis are provided to verify the effectiveness of the proposed technique. To date, no CGH algorithm is available to achieve experimental results comparable to those presented in this work using merely a single pair of commercial SLM and light source. We believe that our approach offers a novel and feasible solution to address the long-standing issues in the holography community and potentially makes holographic displays a viable technology, particularly for emerging VR and AR applications.

The current prototype still utilizes digitally ways to shift the hologram according to the learned motion trajectory, and the reconstructions are obtained offline. However, it is not nontrivial to extend the approach to physical domain: commercial piezoelectric chip can achieve resonant frequency exceeding 60 kHz and maximal displacement beyond 15 $\mu$m, which significantly surpass the requirement of the real-time displays and the displacement in each direction of the optimized trajectory. In the future, we aim to incorporate piezoelectric chips to mount and physically displace the SLM for real-time and dynamic 3D displays to demonstrate the full potential of the proposed technology.

The hologram generation of the proposed approach is currently limited by the computational time, since each hologram needs to be individually computed in Motion Holography. For example, to generate a 1080p hologram using the proposed method, 2000 iterations of optimization are required which takes approximately 8 minutes in total. Although the iterative optimization of the proposed holography can be conducted on the target plane (Fig. S7), which allows the optimization time to align with that of the traditional holography, it still falls short of the real-time hologram generation requirements. To address this issue, we plan to implement the neural network, which appears to be a possible candidate for the real-time hologram generation (*10, 12*), to model Motion Holography for practical applications.

# Materials and Methods

## Simulation configurations

We use DeepFocus (*47*) to synthesize target focal stacks with natural focus cues. In simulation, seven target images are rendered as a volume, and their distances are measured to be -3, -2, -1, 0, 1, 2, and 3 mm away from the conjugate plane of the SLM. The wavelengths of light sources are 680 nm, 520 nm, and 450 nm, which correspond to red, green, and blue laser diodes, respectively. The resolution of the hologram is 2048×1152 with a 3.74 $\mu$m pixel pitch, while the resolution of the target image of interest is 1920×1080. All simulations are optimized and tested in PyTorch using an NVIDIA GeForce RTX 3090 GPU Card.

We randomly select 50 test images from the DIV2K validation set (38), the Big Buck Bunny video, which comes from www.bigbuckbunny.org (2008, Blender Foundation) under the Creative Commons Attribution 3.0 license (https://creativecommons.org/licenses/by/3.0/), and the MPI Sintel dataset (39) to evaluate the generalization capability of the proposed method. To assess the quality of the reconstructed images, we employ widely used evaluation metrics such as PSNR and SSIM (40), which provide quantitative measures to evaluate the fidelity and similarity of the reconstructed images compared with the ground truth.

## PPO model architecture and training configurations

Within the RL framework, we restrict the movement directions $d \in \{\text{up}, \text{down}, \text{right}, \text{left}\}$ and the pixel value per move $p \in \{1,2,3\}$, which means that the number of valid actions is 12 in this case. Both the actor and critic networks are three fully-connected layers. Specifically, for the actor network, the first layer consists of 3 input nodes and 64 output nodes, and the second layer maintains 64 nodes for both input and output. The last layer is a fully-connected layer with 64 input nodes and provides a single output for each valid action. Both first and second layers use the ReLU activation function, while the last layer utilizes the Softmax activation function.

For the critic MLP, the first layer and the second layer are identical to the first and second layers of the action network. The last layer has 64 input nodes and 1 output nodes. To train the deep agent, the learning rate is set to be $1 \times 10^{-4}$ for actor MLP and $3 \times 10^{-4}$ for critic MLP, respectively, both of which are optimized by using the Adam optimizer with $\beta_1 = 0.9$ and $\beta_2 = 0.999$ to update network parameters. The total reward is calculated after taking 8 steps, and the environment is then reset for the next iteration. We collect 160 pairs of data at a time in the buffer to train the actor and critic networks with 200 epochs. The clip parameter is 0.2 and the discount factor is 0.99. The complete set of hyper-parameters used for training deep agent is listed in the supplementary materials. The RL framework is trained on the NVIDIA A100 Tensor Core GPU.

## Experimental setup

The optical setup is shown in Fig. 5, which includes a HOLOEYE GAEA-2 phase-only liquid crystal on silicon with a resolution of 3840×2160 pixels and a pixel pitch of 3.74 $\mu$m. This SLM has a refresh rate of 60 Hz (monochrome) and 8-bit depth. The laser used is a single-mode fiber-coupled module, which consists of three laser diodes optically aligned at wavelengths of 680 nm, 520 nm, and 450 nm. Specifically, to avoid chromatic artifacts, the emission intensities of the red, green, and blue channels of the multi-wavelength laser diodes are set to 60 %, 20 %, and 10 % of their respective maximum values of 40 mW in our optical setup. It should be noted that in the implementation, the color images are captured by individually exposing each wavelength and then merged in post-processing. The collimating lens (CL) is an achromatic doublet with a focal length of 100 mm and we utilize the beam expander (BE) to ensure that the laser incident on the SLM is as uniform as possible. The holographic setup includes two polarizers, where the first polarization filter is used to control the intensity of the laser and the second polarization filter is used to match the polarization direction of the SLM. All images are captured by a CCD sensor (BFS-U3-31S4M-C, FLIR, US) with a resolution of 2048×1536 pixels and a pixel pitch of 3.45 $\mu$m. We employ nearest-neighbor interpolation to double the size of the hologram, effectively alleviating the crosstalk that occurs between neighboring pixels on the SLM. Therefore, to ensure that the camera captures the complete reconstructed plane, we set the resolution of the hologram to 1024×576 and the resolution of the reconstructed image to 960×540 in the experiment. We provide a 4F system, where the first achromatic lens has a focal length of 60 mm, and the second achromatic lens has a focal length of 50mm. An iris is positioned at the Fourier plane to block excessive light diffracted by the grating structure and higher-order diffractions.

**System calibration**

We employ the CITL optimization (*8, 25*), which is used to correct the mismatch between the optical forward propagation in the real world and ideal wave propagation in the simulation, to further enhance the experimental results, as shown in Fig. 6. A detailed explanation of the calibration procedure is given in the following paragraphs.

We adopt the approach of Choi et al. (*11*) to design a real-world forward model where unknown parameters are learned from a dataset of experimentally captured SLM-image pairs. Specifically, we use $\text{CNN}_{\text{SLM}}$ in the SLM plane to assist the ASM model to address the non-uniform intensity distribution of the laser source and nonlinear mapping from voltage to phase delay of the SLM. Additionally, a $\text{CNN}_{\text{Target}}$ in the target plane is utilized to handle optical aberrations and camera sampling issues in the holographic system. We further introduce $a_{src}$ and $\phi_{src}$ to learn the content-independent spatial variations in amplitude and phase of the incident source field at the SLM plane. In conclusion, our learnable propagation model can be expressed as:

$$f_{\text{model}}(\phi, z^{\{j\}}) = \text{CNN}_{\text{Target}}\left(f_{\text{ASM}}\left(\text{CNN}_{\text{SLM}}(a_{\text{src}}\, e^{i\phi_{\text{src}}}\, e^{i\phi}), z^{\{j\}}\right)\right). \quad (11)$$

To train $f_{\text{model}}\{\cdot\}$, we randomly generate 2400 holograms using traditional and proposed methods, and then capture the intensity of holograms in 7 target depth planes for each color channel. During the hologram generation process, we randomize the learning rate, the optimization iteration and the initial random phase distribution to ensure the generalization of the learned propagation model (*11*). Both $\text{CNN}_{\text{SLM}}$ and $\text{CNN}_{\text{Target}}$ are implemented using U-Net (*50*) with two input and output channels for the real and imaginary components of the fields. For training hyper-parameters, the batch size is 1 and the initialization learning rate is $3 \times 10^{-4}$. We train the network for 100 epochs, using the AdamW (*51*) optimizer with momentum of (0.9, 0.999), and then use the cosine decay strategy to decrease the learning rate. All experiments are trained and tested using an NVIDIA GeForce RTX 3090 GPU Card.

## Acknowledgments
The authors wish to thank Jidong Jia for his invaluable assistance with reinforcement learning training.

## Funding
This work was supported by National Key Research and Development Program of China (2021YFB2802100, 2023YFB3611501).

## Author contributions
Z.D. and Y.L. conceived the idea and wrote the manuscript. Z.D. developed the software/hardware system, captured the results, and created the figures. Y.L. , Y.L. , and Y.S. supervised the project. All authors reviewed and contributed to the manuscript.

## Competing interests
The authors declare that they have no competing interests.

## Data and materials availability
All data needed to evaluate the conclusions in the paper are present in the paper and/or the Supplementary Materials.

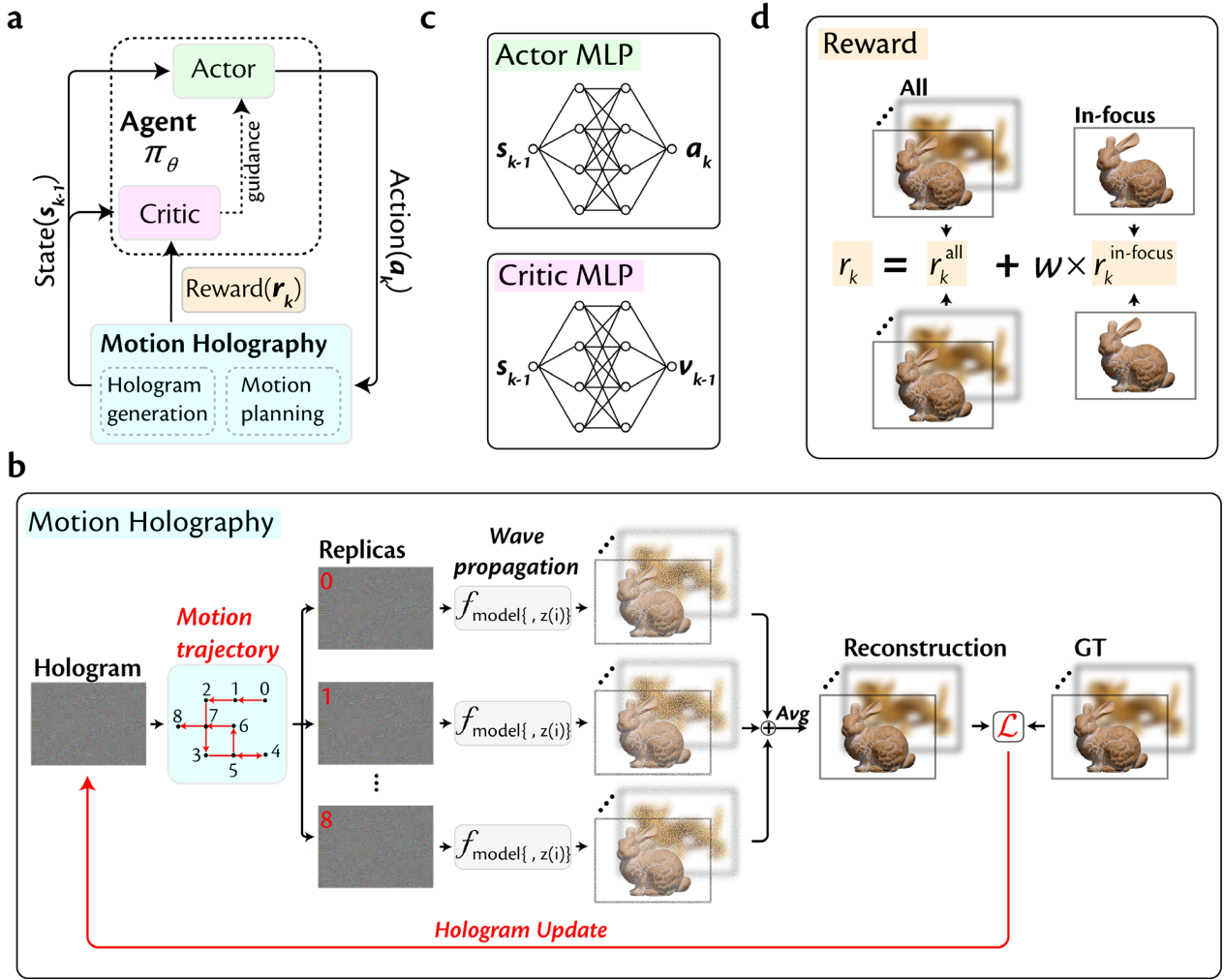

**Fig. 1. Overview of reinforcement learning-empowered jointly-optimized hologram generation and motion planning.** (a) The deep agent $\pi_\theta$ interacts with the environment, i.e. Motion Holography, in a sequence of actions $a$, observations $s$ and rewards $r$. The goal of the deep agent is to strategically choose actions to maximize the expected cumulative reward and generate an optimized motion planning. (b) The motion hologram is moved to specific positions according to the updated motion trajectory to generate the copies of the hologram. Each copy of the hologram is propagated to reconstruct the 3D scene with specially-designed "motion speckles". All the reconstructed 3D scenes are superimposed and averaged to obtain the final reconstruction scene. (c) The actor network learns the policy to select actions, while the critic network provides feedback on the quality of these actions, enabling the actor network to improve its policy based on the received rewards. (d) The immediate reward $r_k$ is composed of two parts, $r_k^{\text{all}}$ and $r_k^{\text{in-focus}}$.

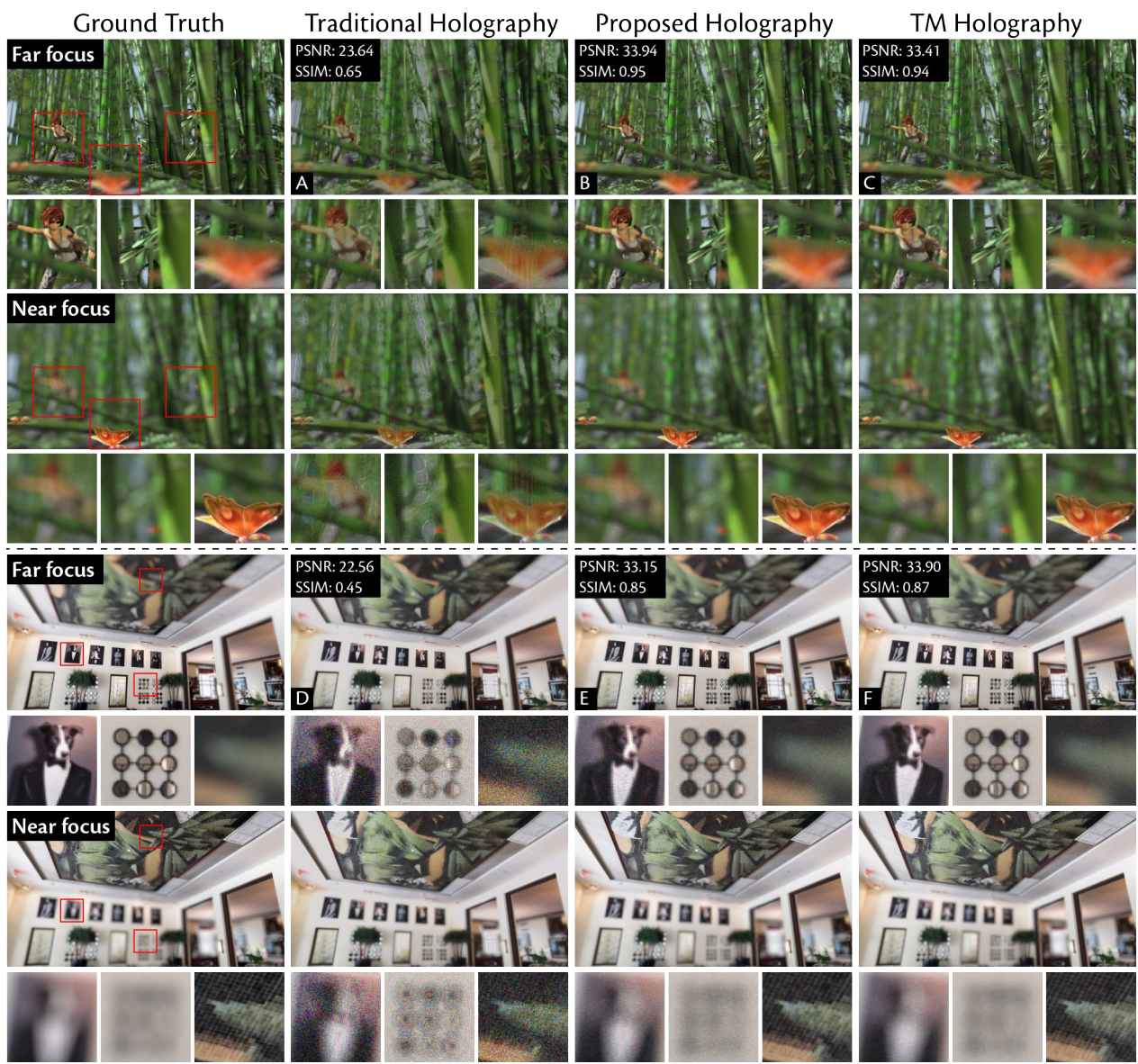

**Fig 2. Simulation results.** Comparison of 3D holograms synthesized by using different methods including traditional holography, proposed holography, and time-multiplexed holography. In both examples, traditional holography induces significant speckle noise, unnatural ringing artifacts, and severe chromatic aberration. In contrast, the proposed holography can reconstruct high-quality and speckle-free 3D scenes with high-resolution in-focus regions and realistic out-of-focus areas. Furthermore, we demonstrate that the proposed holography achieves comparable visual results to that of the state-of-the-art time-multiplexed holography. Image Credits: Daniel J Butler, University of Washington, and Eirikur Agustsson and Radu Timofte, ETH Zurich.

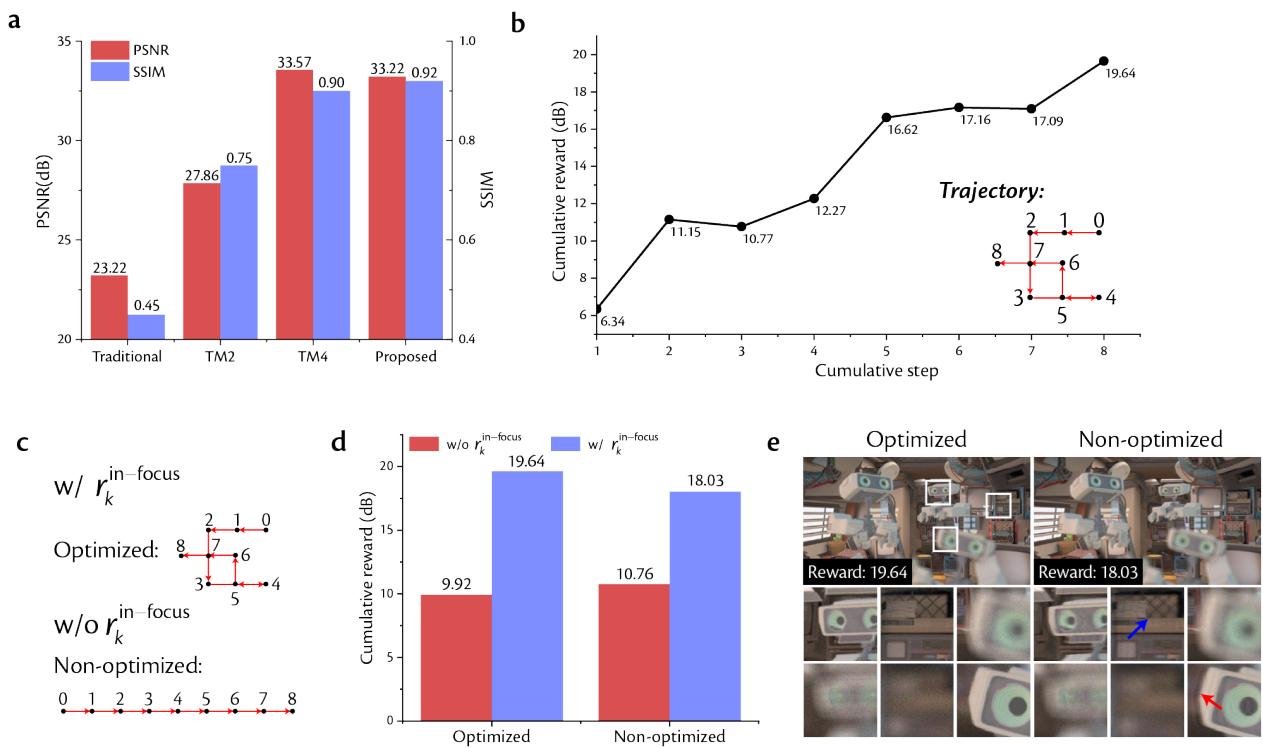

**Fig. 3 Model analysis. (a)** Quantitative comparison between the PSNR and SSIM obtained by traditional holography, two-frame time-multiplexed holography (TM2), four-frame time-multiplexed holography (TM4), and the proposed holography. **(b)** The cumulative rewards obtained at each step along the optimal trajectory. **(c)** The trajectories obtained through reinforcement learning optimization with and without $r_k^{\text{in-focus}}$, respectively, labeled as Optimized and Non-optimized. **(d)** When incorporating $r_k^{\text{in-focus}}$, the proposed holography achieves a higher total reward using optimized trajectory compared with using non-optimized trajectory. **(e)** In the visual effects, the 3D reconstructed scene based on non-optimized trajectory exhibits undesirable motion blur and introduces ringing artifacts. The proposed method successfully overcomes these challenges using optimized trajectory. Images Credits: Lei Xiao, Meta.

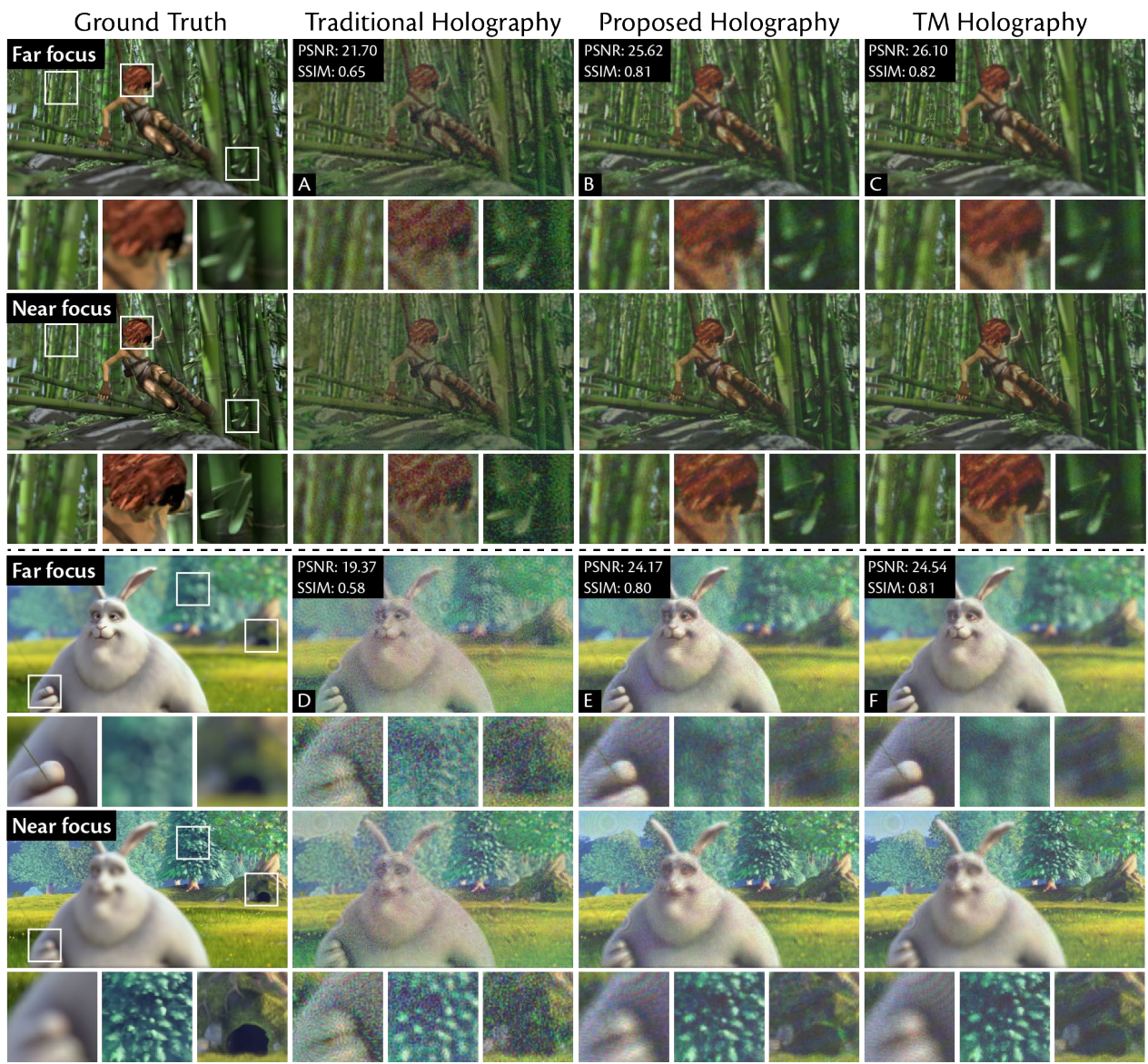

**Fig. 4 Experimental results.** Comparison of 3D holograms synthesized using several different methods in experiment, including traditional holography, proposed holography, and time-multiplexed holography. Traditional holography suffers from speckle noise and low contrast in the reconstructed 3D scene due to the insufficient spatial bandwidth product of a single SLM. Time-multiplexed holography utilizes multiple time-incoherent holograms to reconstruct high-quality and photorealistic 3D scenes, however, it requires high-speed SLM. Compared with these holographic techniques, proposed holography benefits from breaking spatial coherence using learnable motion trajectory to achieve the speckle-free and photorealistic 3D images for both in-focus and out-of-focus regions. Images Credits: Daniel J Butler, University of Washington, and Big Buck Bunny, Blender Institute.

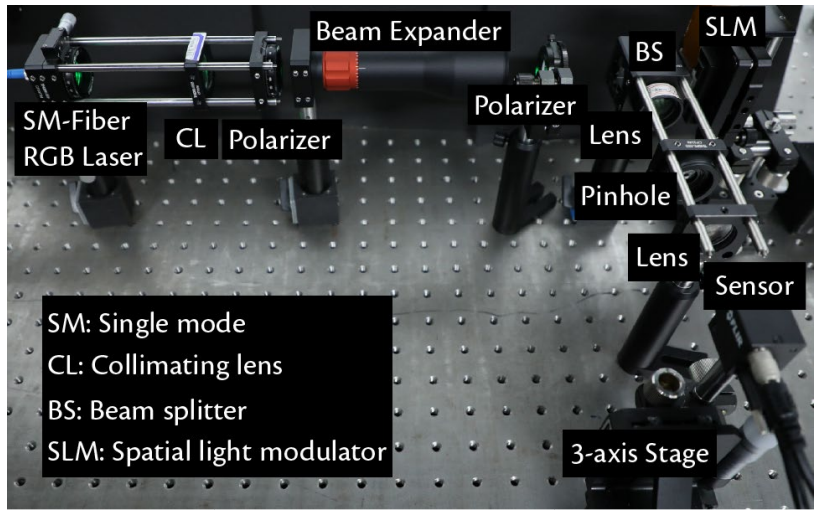

**Fig. 5 Illustration of our holographic display setup.** An RGB laser module is coupled to a single mode (SM) fiber to illuminate our phase-only SLM through a collimating lens, and two polarizers. We provide a 4F system, where a pinhole is positioned at the Fourier plane to block excessive light diffracted by the grating structure and higher-order diffractions. The sensor is used to capture multi-plane images by moving the 3-axis stage.

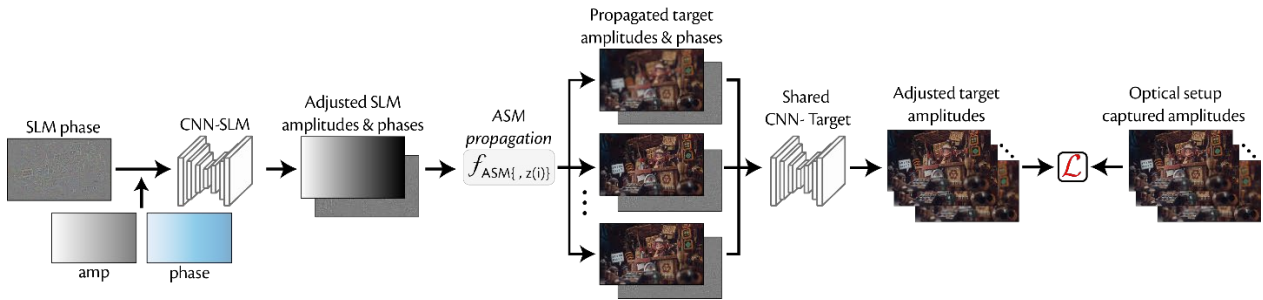

**Fig. 6 Illustration of the camera-in-the-loop (CITL) model.** The phase pattern displayed by the SLM is adjusted by learnable amplitude and phase and the is processed by a CNN. The resulting complex-valued wave field is propagated to all target planes using a conventional ASM wave propagation operator. The wave fields are processed again by a shared CNN at each target plane. The loss function is calculated as the mean squared error (MSE) between the adjusted target amplitude and the amplitudes captured at different planes using the optical system. Images Credits: Alex Treviño.

# Supplementary Materials

## 1. Jointly optimize hologram generation and motion planning with deep reinforcement learning

In this work, we develop a deep reinforcement learning framework to jointly optimize hologram generation and motion planning to reconstruct the optimal photorealistic 3D contents in an end-to-end manner. We expect the deep agent $\pi_\theta$ interacts with the Motion Holography by strategically choosing actions to maximize the quality of reconstructed images. To do end, we firstly define the action, state, environment, and reward in our task, as listed in Table S1.

The motion trajectory $\tau$ is defined as the actions and states obtained from the continuous interaction between an agent and an environment. Since the variation of actions output by the deep agent $\pi_\theta$ in each interaction, the probability of a complete trajectory can be defined as follows:

$$p_\theta(\tau) = p(s_1)\prod_{t=1}^{T} p_\theta(a_t \mid s_t)p(s_{t+1} \mid s_t, a_t), \text{(S1)}$$

the reward obtained from a sequence $\tau$ is the accumulation of $r$ obtained at each stage, referred to as $R(\tau)$. Therefore, the expected reward that can be obtained is:

$$J(\theta) = \bar{R}(\theta) = \sum_\tau R(\tau)p_\theta(\tau) = \mathbb{E}_{\tau \sim p_\theta(\tau)}[R(\tau)]. \text{(S2)}$$

To maximize the expected reward, we employ the Policy Gradient (PG) method, which applies the gradient ascent algorithm to gradient estimation and utilizes the Monte Carlo (MC) method to estimate the gradient:

$$\nabla_\theta J(\theta) = \mathbb{E}_{\tau \sim p_\theta(\tau)}[R(\tau)\nabla \log p_\theta(\tau)] \approx \frac{1}{N}\sum_{n=1}^{N} R(\tau^n)\nabla \log p_\theta(\tau^n). \text{(S3)}$$

The entire optimization process can be unfolded as follows: the agent $\pi_\theta$ continuously interacts with the environment, collecting actions, states, and corresponding rewards. Next, the parameters of agent are updated using the gradient ascent method described in Eq. S3. Subsequently, new data is collected again under the updated policy, followed by a further update of the parameters. This iterative process is repeated until the optimal trajectory is achieved. Generally, the PG method updates the policy by increasing the likelihood of high-reward actions and decreasing the probability of low-reward actions. However, when all action rewards are positive, the outcomes obtained using the PG method may lead to a significant variance. To address this issue, we adopt the actor-critic algorithm, which introduces a state-action-dependent advantage function $A^\pi(s, a)$, as follows:

$$A^\pi(s, a) = Q^\pi(s, a) - V^\pi(s), \text{(S4)}$$

where $Q^\pi(s,a) = \sum_t \mathbb{E}_{\pi_\theta}[R(s_t, a_t) \mid s, a]$ and $V^\pi(s) = \sum_t \mathbb{E}_{\pi_\theta}[R(s_t, a_t) \mid s]$. The Q-function, $Q^\pi(s,a)$, represents the expected total reward for a given series of states $s$ and corresponding actions $a$, while the value function, $V^\pi(s)$, which is used to evaluate the policy $\pi$, can be implemented using a critic network. Therefore, the advantage function $A^\pi(s,a)$ quantifies the relative benefit of an action over others in the same state.

The naïve PG method is an on-policy strategy, which means that every parameter update requires re-sampling and leads to low data utilization and long convergence times. To overcome this challenge, we introduce the proximal policy optimization (PPO), which employs importance sampling to estimate the expected values from samples collected under an old policy $\pi_{\theta'}$, to enhance the convergence rate of the PG method. In this case, the expected reward can be reformulated as:

$$J(\theta) = \mathbb{E}_{\tau \sim p_{\theta'}(\tau)}\left[\frac{p_\theta}{p_{\theta'}} A\right] = \sum_{t=1}^{T} \frac{p_\theta(a_t|s_t)}{p_{\theta'}(a_t|s_t)} A_t^{\theta'}(s_t, a_t). \text{(S5)}$$

As the new policy $\pi_\theta$ is trained, the two policies $\pi_\theta$ and $\pi_{\theta'}$ will diverge in terms of the action outputs for the identical state, which will lead to increased estimation variance. To mitigate this issue, the parameters of old policy $\pi_{\theta'}$ is periodically updated to align the parameters of new policy.

The objective function of PPO can be denoted as follows:

$$\mathcal{L}(\theta) = \sum_{(s_t, a_t)} \left[\min\left(r_t(\theta) A_t^{\theta'}, \text{clip}(r_t(\theta), 1-\epsilon, 1+\epsilon) A_t^{\theta'}\right)\right], \text{(S6)}$$

where $r_t(\theta) = \frac{p_\theta(a_t|s_t)}{p_{\theta'}(a_t|s_t)}$. PPO incorporates two safeguards into the objective function to prevent instability in policy updates due to overly large update steps and policy ratios. The first insurance is a restriction on policy ratio $r_t(\theta) \in [1-\epsilon, 1+\epsilon]$. This avoids the algorithm from excessively favoring actions with positive advantages or hastily discarding actions with negative implications based on limited samples. The second safeguard is the use of a min function, which ensures that the surrogate objective function serves a lower bound for the unclipped objective and mitigates the bias against actions with negative advantages.

The framework is implemented using Python and PyTorch, and the pseudocode for the hologram generation and motion planning optimization is presented in Algorithm S1. We further provide the values and descriptions of all hyper-parameters in Table S2.

## 2. Supplementary simulation results

**Synthesizing the motion speckle pattern.** In this work, we explored the use of motion to synthesize unique speckle patterns and achieve photorealistic and speckle-free 3D display via effective speckle suppression. Figure S1 compares the speckle patterns generated in the green channel by both the traditional and proposed methods. The 3D scenes reconstructed using the traditional method exhibit low resolution and disordered speckle noise. In contrast, the proposed single hologram reconstruction method yields more regular and structured speckle patterns, due to the jointly optimization with motion planning. Despite a lower PSNR of 17.2 dB for the scene reconstructed using our method, it clearly offers multi-plane images with natural focus cues, which is unachievable with the traditional approach. By averaging the holograms of the proposed method along the motion trajectory, we can reconstruct high-fidelity 3D scenes with higher-resolution and photorealistic contents (Figure S1, Column 4). A further comparison of holograms from both the traditional and proposed holography is present in Figure S1, Line 5.

**Averaging along with motion trajectory.** To demonstrate the effectiveness of the proposed hologram generation algorithm, we compared reconstructed images using four approaches: the traditional hologram, the traditional hologram along the proposed motion trajectory, the proposed hologram, and the proposed hologram along the motion trajectory. The visual results are shown in Figure S2. It is apparent that the speckles generated by our method exhibit a structured pattern and the in-focus and out-of-focus effects are clearly visible (Figure S2, Column 3). After averaging the holograms generated by both methods along the jointly optimized motion trajectory, it can be seen that while the traditional holography partially suppresses speckles, the reconstructed images suffer from lower contrast and resolution due to insufficient optimization (Figure S2, red arrow). Conversely, the proposed holography effectively depicts high-fidelity and photorealistic 3D scenes through the joint optimization of hologram generation and motion planning. We also present the corresponding holograms in Figure S2.

**Learnable trajectories and rewards at various step sizes.** We conducted ablation study to investigate the learned motion trajectories and their corresponding cumulative rewards at various step sizes, are shown in Figure S3. Initially, with an increase in total step size, there is a continuous increase in cumulative reward. However, the cumulative reward slightly decreases when the step size exceeds nine. We further present the visual effects of reconstructed images at different step sizes in Figure S4. It can be clearly observed that the proposed holography can faithfully reproduce high-quality and photorealistic 3D scenes at a total step size of five. Considering practical deployment, where longer step sizes is required faster displacement stages and improved mechanical stability, we selected the results at a step size of eight for detailed presentation in the main paper. It is curcial to note that even at smaller step sizes, our method still outperforms traditional holography in reconstructing 3D scenes with natural focus cues.

**Ablation study in $r_{\text{infocus}}$.** To further demonstrate the importance of $r_{\text{infocus}}$, we presented the reconstructed results of the USAF-1951 resolution chart in the green channel using the traditional method, and proposed method based on both trajectories, as illustrated in Figure S6. Due to the binary pixel values of the USAF-1951, we only rendered three planes in the simulation to simplify the optimization problem. The resolution chart is projected directly onto the first plane, with the remaining two planes are synthesized by DeepFocus to generate realistic defocus images. We observe that although the proposed holography using both trajectories effectively suppress speckle noise compared to the traditional holography, the proposed holography using Trajectory 1 introduces motion blur and lower resolution contents in the reconstructed image due to its unidirectional movement (Figure S6, red arrow). In comparison, the proposed holography using Trajectory 2 achieves the higher-resolution reconstructed image.

**Comparing the reconstructed results of motion trajectory between the SLM plane and the target plane.** Here, we analyzed the reconstructed results using the proposed method on both the SLM plane and the target plane, as depicted in Figure S7. In Figure S7 A, both the optimization and reconstruction of the hologram occur on the SLM plane, whereas both processes are conducted on the target plane in Figure S7 D. It is evident that our methods both achieve superior image quality regardless of the plane in which the hologram is moved. This is due to the short propagation distance of near-eye holography. For example, in our paper, the propagation distance ranges from -3 mm to 3 mm. Therefore, the holograms in the SLM plane and target plane can be considered to exist within the same domain. To validate our hypothesis, we conducted cross-validation experiments. Figure S7 B illustrates the reconstruction results, with the optimization on the SLM plane and the reconstruction on the target plane, while Figure S7 C shows the reconstruction results the optimization on the target plane and the reconstruction on the SLM plane. Both qualitative and quantitative analyses reveal minimal differences in the image quality, regardless of whether optimization and reconstruction occur on the same plane. Therefore, we adopt optimization on the target plane during both PPO training and CITL training to significantly reduce computation time and GPU memory usage. Furthermore, the proposed method using this optimization approach aligns the optimization time with traditional method and significantly reduces it compared to time-multiplexed holography.

**Additional simulated results.** Additional comparison results of traditional, proposed, and time-multiplexed holography are presented in Figure S8 and S9. We also show the corresponding holograms and suggest readers to zoom in the images for a clearer comparison of different methods.

## 3. Supplementary experimental results

**System calibration.** In this work, we have demonstrated in simulations that Motion Hologram is a promising technology. However, replicating these results requires accurate wave propagation in the holographic system. While the angular spectrum method (ASM) performs well in simulations for forward propagation, but it inadequately represents real-world optical systems. This discrepancy arises from real-world system imperfections such as optical aberrations, imperfect beam collimation, nonlinear SLM responses, and limited SLM diffraction efficiency, none of which can be accurately modeled by the ASM. Therefore, the mismatch between the simulation and experimental propagation models leads to suboptimal experimental results.

To address this issue, we employed the method proposed by Choi et al. to develop a differentiable camera-calibrated model for the wave propagation in multi-plane holographic displays, as shown in Figure 6 of the main paper. Both $CNN_{slm}$ and $CNN_{target}$ are implemented using the classic U-Net architecture with two input and output channels to handle the real and imaginary components of the wave fields. The $CNN_{slm}$ model incorporates skip connections and employs four consecutive down-sampling operations using strided convolutions and four matching up-sampling stages with transposed convolutions. Following the input layer, this CNN starts with 32 feature channels which double at each down-sampling stage, reaching up to 256 channels. The $CNN_{target}$ network comprises three down-sampling and up-sampling layers with 16 feature channels after the input and doubling at each down-sampling stage to a maximum of 128 channels. Both networks employ instance normalization LeakyReLU for the down-sampling, and ReLU nonlinearities for the up-sampling. Both $a_{src}$ and $\phi_{src}$ are matrices optimized to match the resolution of the hologram.

For the training dataset, it is crucial to have ample focal stacks with natural focus cues for training the 3D CITL model. A straightforward approach is to utilize rendering engines like Blender and Unity to acquire high-quality focal stack images. However, this method is time-consuming and requires extensive scene resources. To overcome this challenge, we propose an efficient pipeline that generates 3D focal stacks with realistic defocus from RGB images. This framework integrates a monocular depth prediction network with a focal stack synthesis network, as depicted in Figure S10 A. To generate high-quality depth maps, we utilize the state-of-the-art depth prediction network known as Depth Anything, which is a versatile large-scale network specifically designed for accurate estimation of monocular depth maps. The visual results of the predicted color depth map and grayscale depth map based on the input RGB image are shown in Figure S10 B. To render target focal stacks, we employ DeepFocus, an end-to-end CNN that effectively synthesizes defocus blur from RGBD images. We render seven plane images, and showcase three plane images with natural defocus in Figure S10 C.

To train the CITL model, we generate 2400 holograms using both traditional and proposed methods and capture the corresponding intensity of holograms in seven depth planes for each color channel. For hologram generation, source RGB images are randomly selected from high-resolution DIV2K and Flick2K datasets, and the multi-plane images dataset is generated using the proposed focal stack generation pipeline. We randomize the learning rate, the optimization iterations and the initial phase distribution to enhance the generalization of the learned propagation model. Our training data includes 2000 phase patterns and their captured intensity images, while the validation data comprise 400 pairs of phase patterns and images.

Calibration between the hologram and the real-world captured reconstructed image is important for the successful CITL model training. To this end, we implement a planar homography to accurately align the captured images with the ground truth images. We employ a 22×13 white dots binary pattern to calculate the homography. The center-to-center spacing of adjacent dots is 40 pixels. This yields a 960×540 pixel region of interest used for image alignment. Using a 2D stochastic gradient descent CGH algorithm, we generate phase holograms across three color channels and seven planes. Each color channel and plane is required individual calibration, totaling 21 calibrations. A 3×3 homography matrix is computed to transform between the captured and ground truth images. Figure S11 illustrates the calibration results for the first plane of the green channel.

Figure S12 visualizes the trained parameters of our CITL-calibrated model, including the source intensity $a_{src}$, source phase $\phi_{src}$. We present the experimental results using both the ASM and CITL propagation models in Figure S13. It can be clearly seen that the experimental images exhibit chromatic aberrations, noise, and low contrast due to the mismatch between the ASM and the actual optical system propagation. Furthermore, the effectiveness of our proposed method is compromised due to the requirement for precise replication of speckle patterns through motion synthesis. When we employ the CITL model based on the real-world optical system, our method is capable of accurately rendering photorealistic and high-fidelity 3D scenes.

**Additional experimental results.** We present additional comparison experimental results of traditional, proposed, and time-multiplexed holography, as shown in Figure S14, S15, and S16. We also present the corresponding holograms and recommend readers to zoom in the images for better comparison between different methods.

**Table. S1.** Detailed explanation of some parameters in reinforcement learning.

| | |
|---|---|
| Action $a$ | The specific direction ($d$) and pixel ($p$) of each SLM displacement; |
| State $s$ | The current position of the SLM and the number of movement steps ($K$) |
| Environment | Proposed motion hologram generation algorithm |
| Reward $r$ | The improvement in PSNR of the reconstructed image after each movement. |

**Algorithm S1.** Jointly optimized hologram generation and motion trajectory

---

Initialize the policy $\pi_\theta$ with random weight $\theta$
Initialize the old policy $\pi_{\theta'}$ with random weight $\theta'$
Define a Bufffter $B$ to collect the action $a$, state $s$, and reward $r$
**While** *step* $<=$ max _training_setp **do**
  Reset environment and initial state $s$
  **For** $k = 1 \ ... \ K$ **do**
    $a = \pi_{\theta'}(s)$
    $s, r =$ Motion Holography$(a, k)$
    *step* $= + 1$
    $B$.append$(a, s, r)$
    **If** step % training_step $== 0$ **do**
      Optimize surrogate $\mathcal{L}$ wrt $\pi_\theta$, with $N$ epochs
      Reset $B$ and $\pi_{\theta'} \leftarrow \pi_\theta$
  **End**
**End**

**Table S2**. List of hyperparameters and their values

| Hyperparameters | Value | Description |
|---|---|---|
| max_training_timesteps | 40000 | If the step is greater than max_training_timesteps, the training cycle is interrupted |
| max_ep_len | 8 | Total steps in one episode |
| update_timestep | 320 | Update policy every N steps |
| m_epochs | 200 | Update policy for M epochs |
| eps_clip | 0.2 | Clip parameter for PPO |
| gamma | 0.99 | Discount factor |
| lr_actor | 0.001 | Learning rate for actor network |
| lr_critic | 0.003 | Learning rate for critic network 0.001 |
| state_dim | 3 | State space dimension |
| action_dim | 12 | Action space dimension |
| optimizer | Adam | Optimization algorithm for the network training |
| betas | (0.9, 0.999) | Coefficients used by Adam |
| eps | 1e-8 | Term added to the denominator used by Adam |

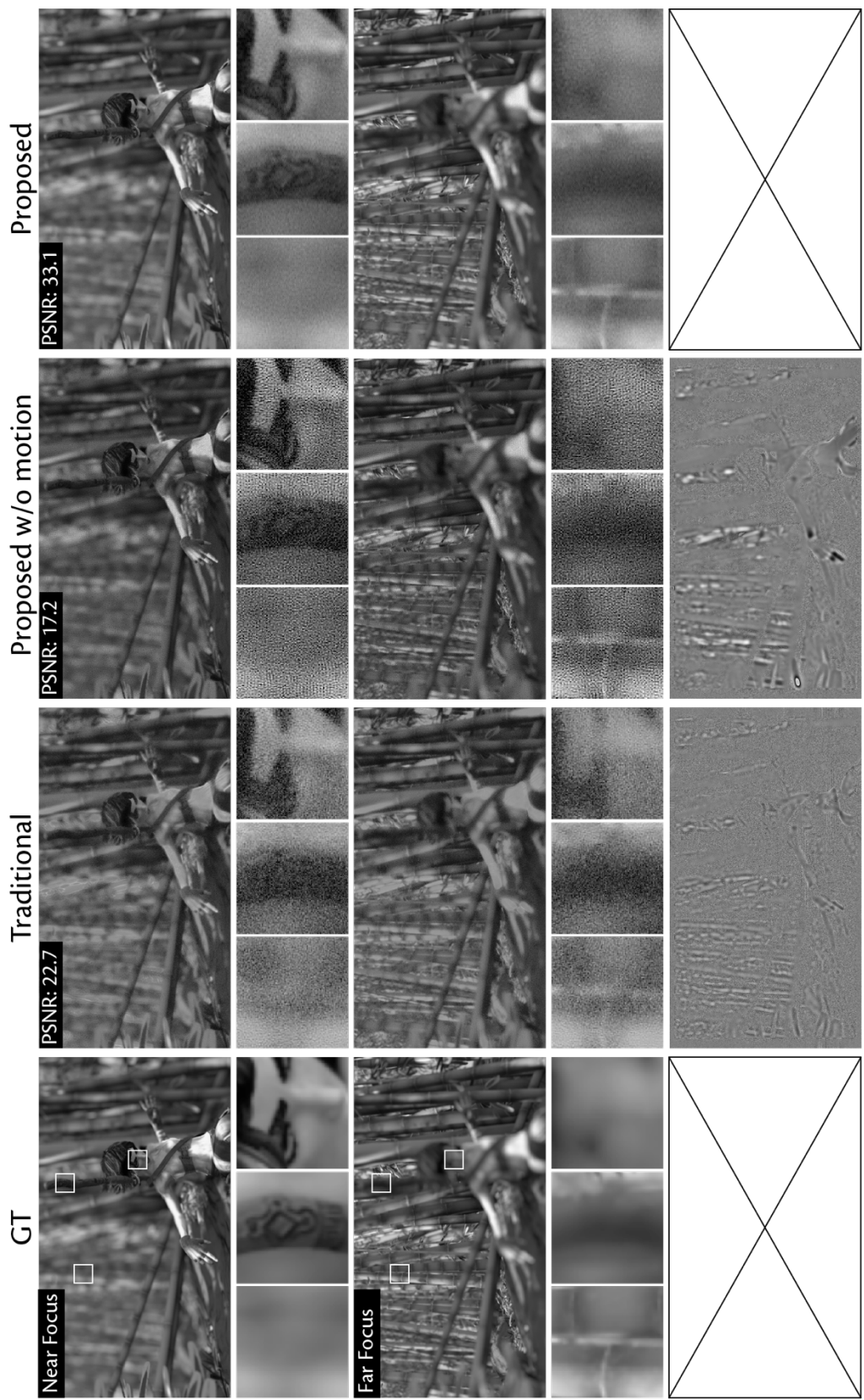

**Figure S1.** Synthesizing the motion speckle pattern in simulation. Image Credits: Daniel J Butler, University of Washington

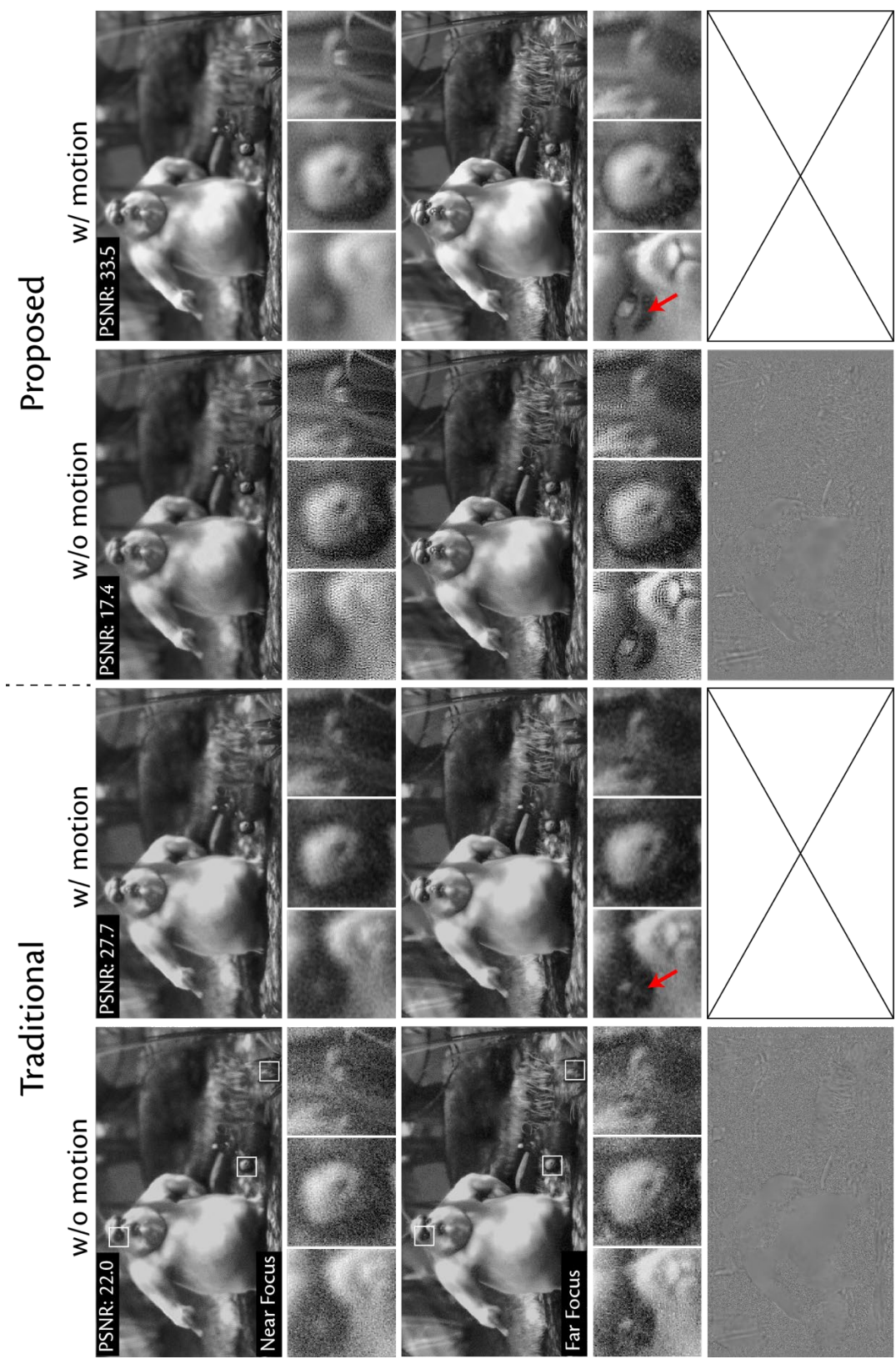

**Figure S2.** Averaging along with motion trajectory. Images Credits: Big Buck Bunny, Blender Institute.

**A**

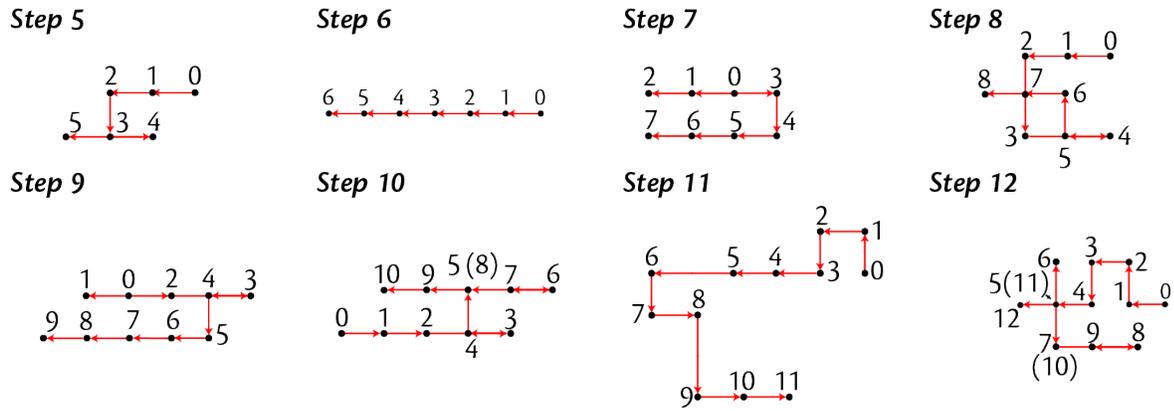

**B**

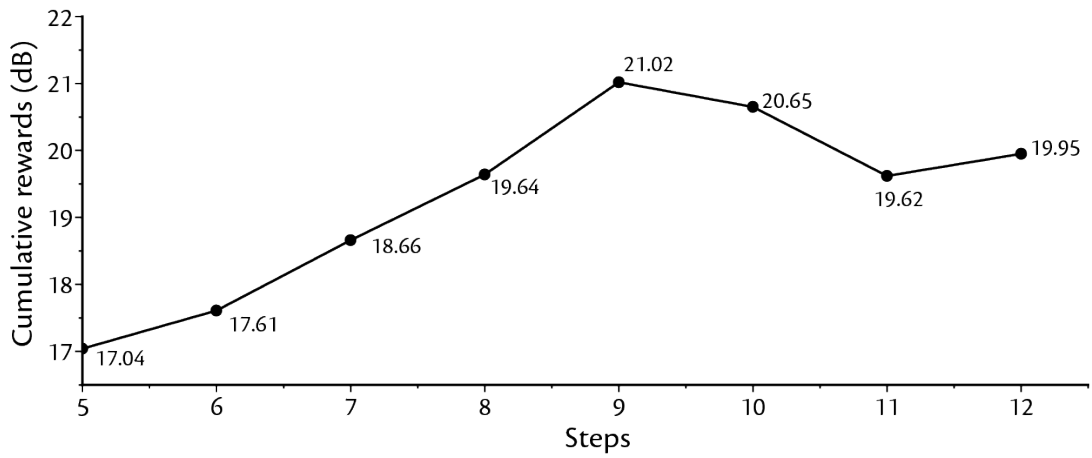

**Figure S3.** Learnable trajectories and corresponding rewards under different step sizes.

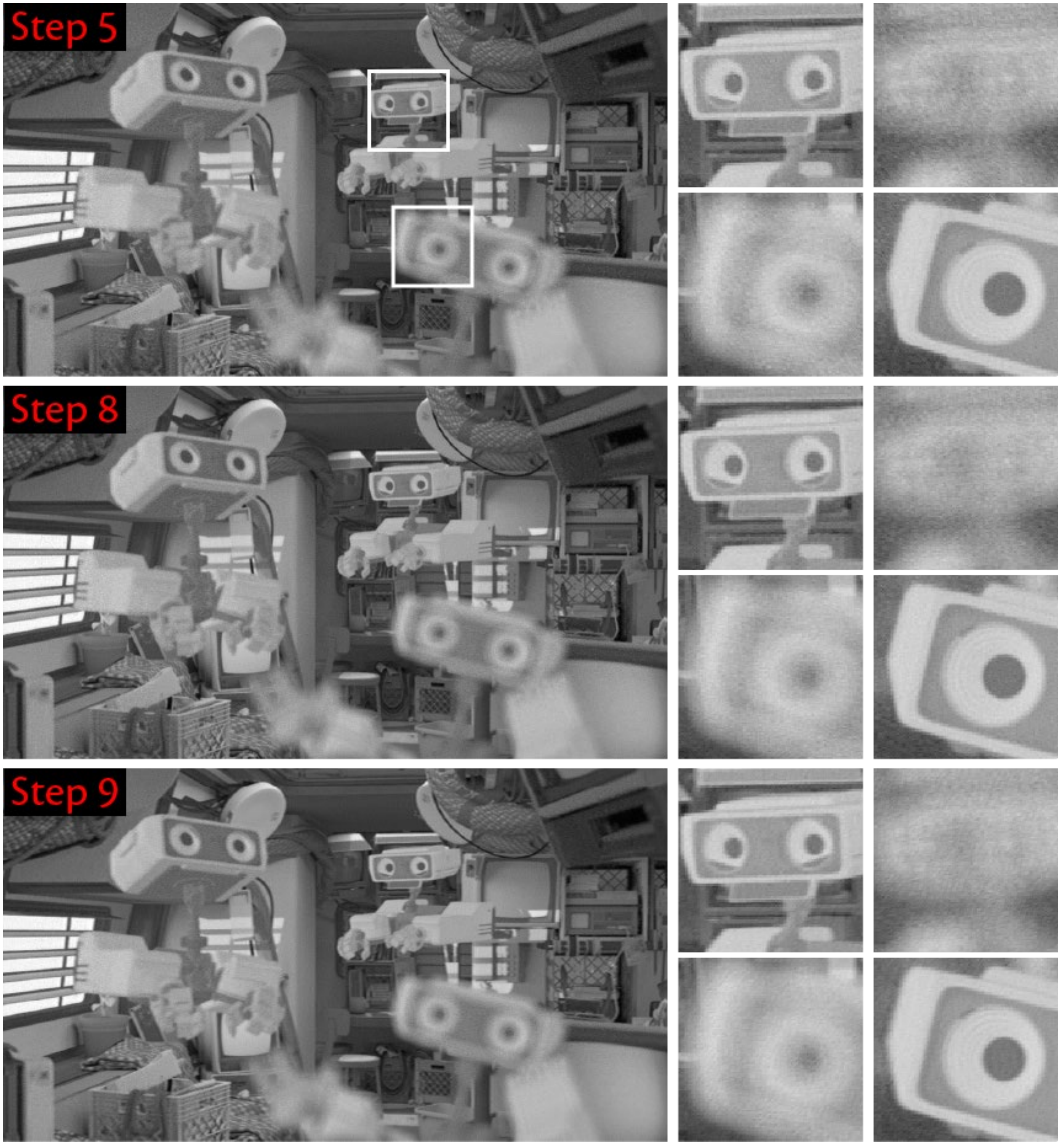

**Figure S4.** The visual results under different step sizes. Images Credits: Lei Xiao, Meta.

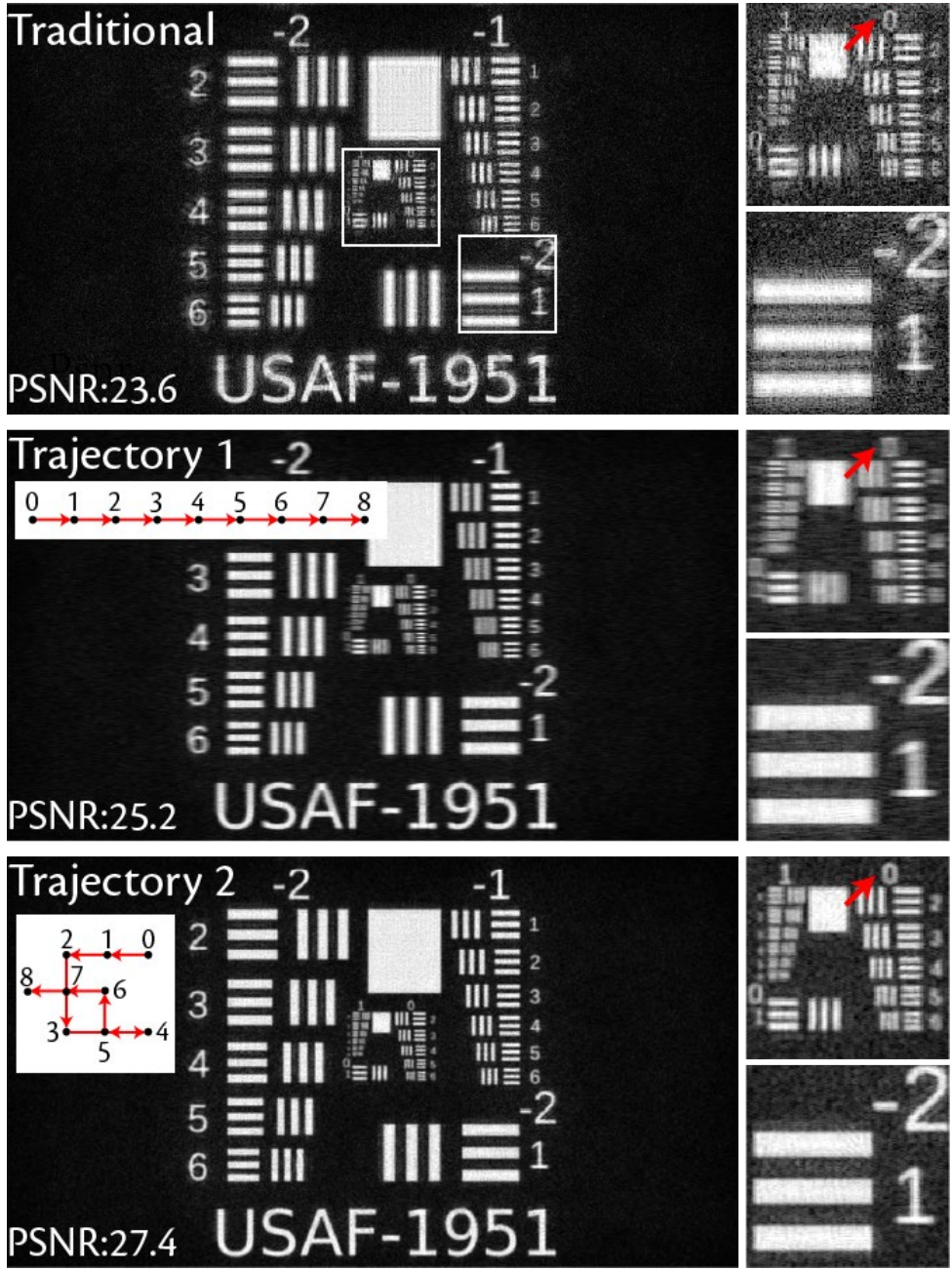

**Figure S6.** Ablation study in $r_{\text{infocus}}$.

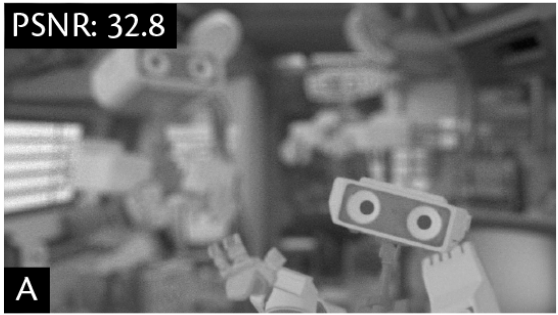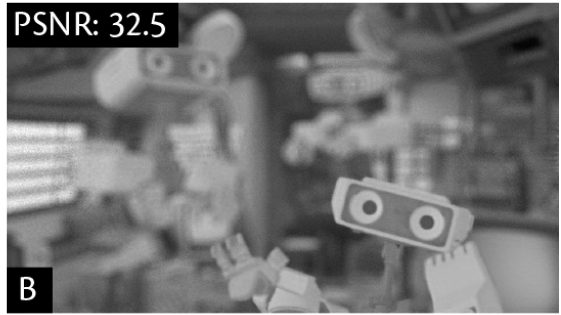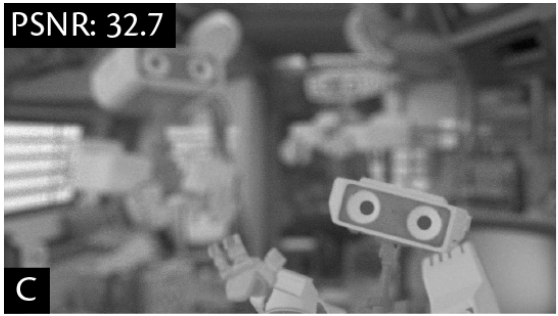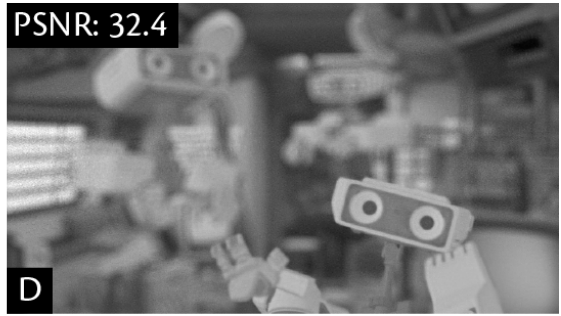

**Figure S7.** Comparing the reconstructed results of motion trajectory between the SLM plane and the target plane. Images Credits: Lei Xiao, Meta.

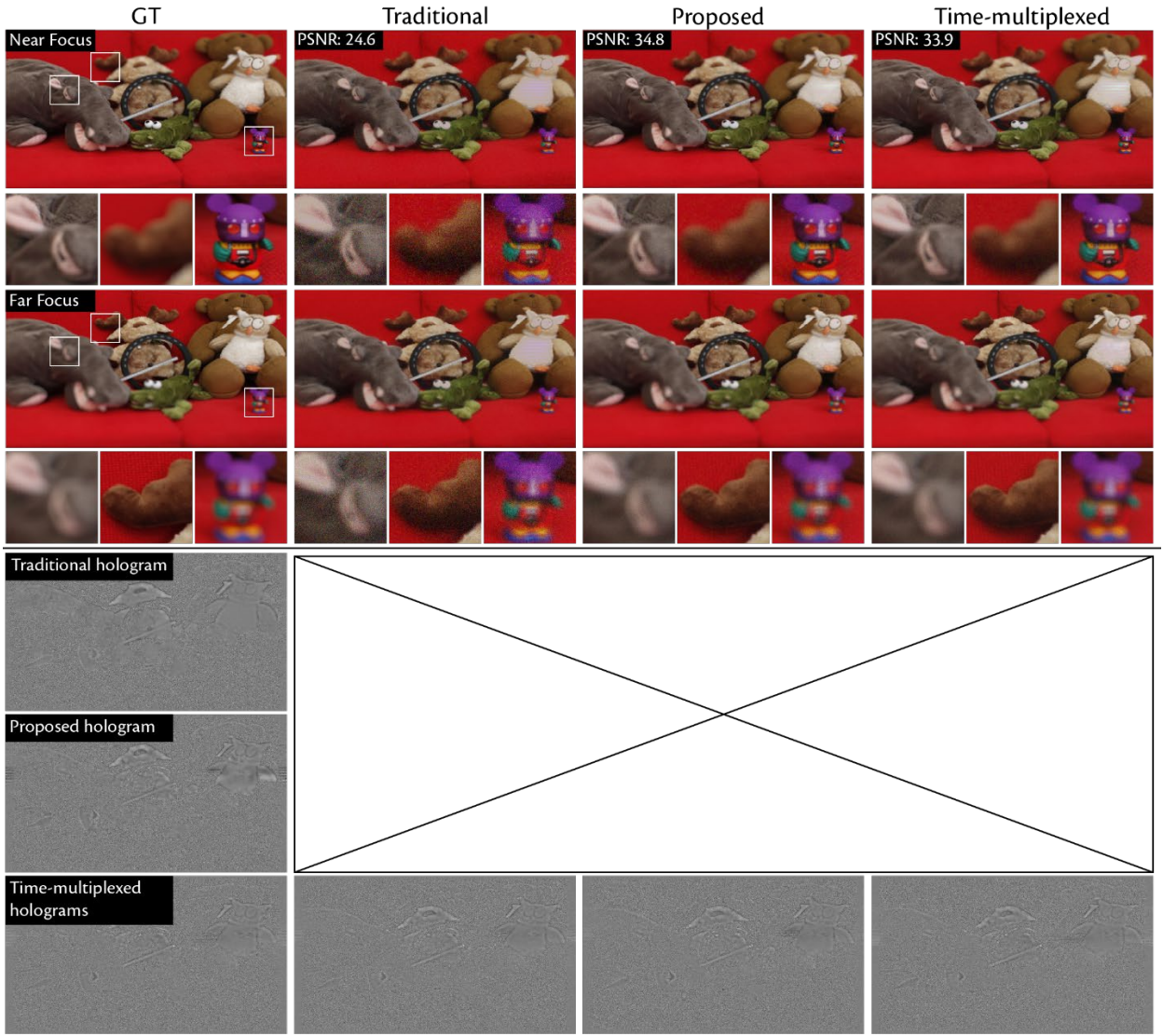

**Figure S8.** Simulation results of images reconstructed with different methods. Images Credits: Changil Kim, ETH Zurich.

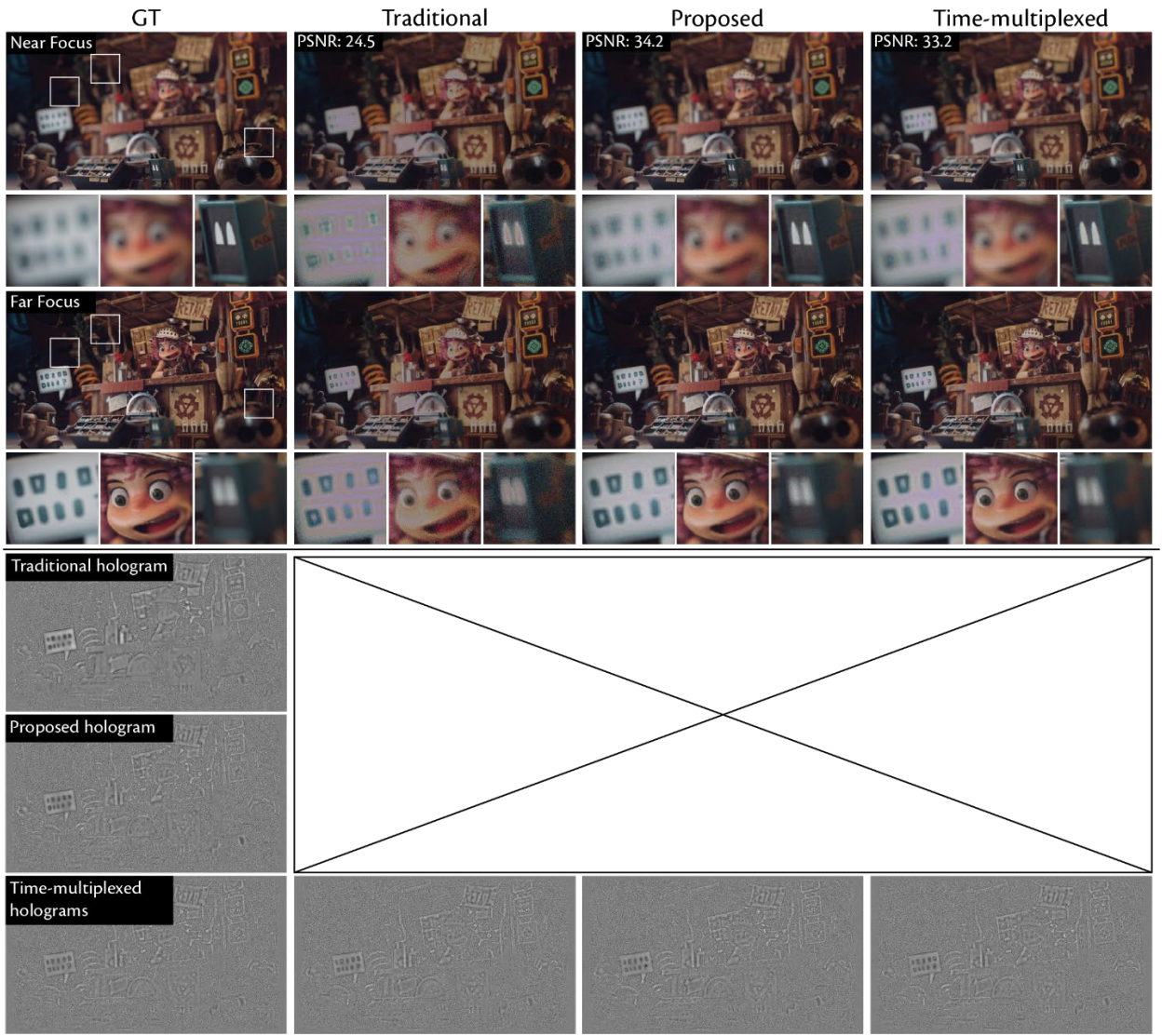

**Figure S9.** Simulation results of images reconstructed with different methods. Images Credits: Alex Treviño.

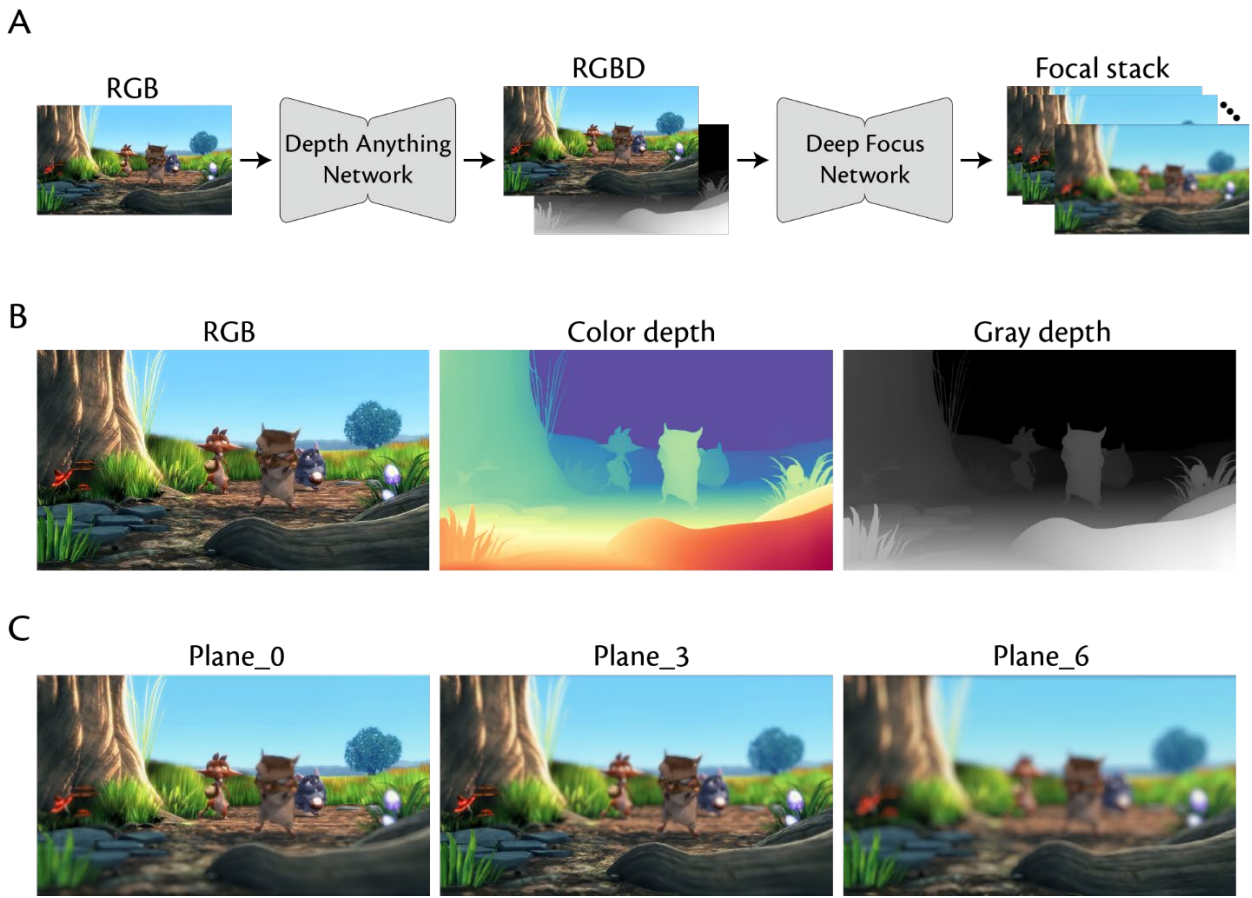

**Figure S10.** Focal stack images generation. Images Credits: Big Buck Bunny, Blender Institute.

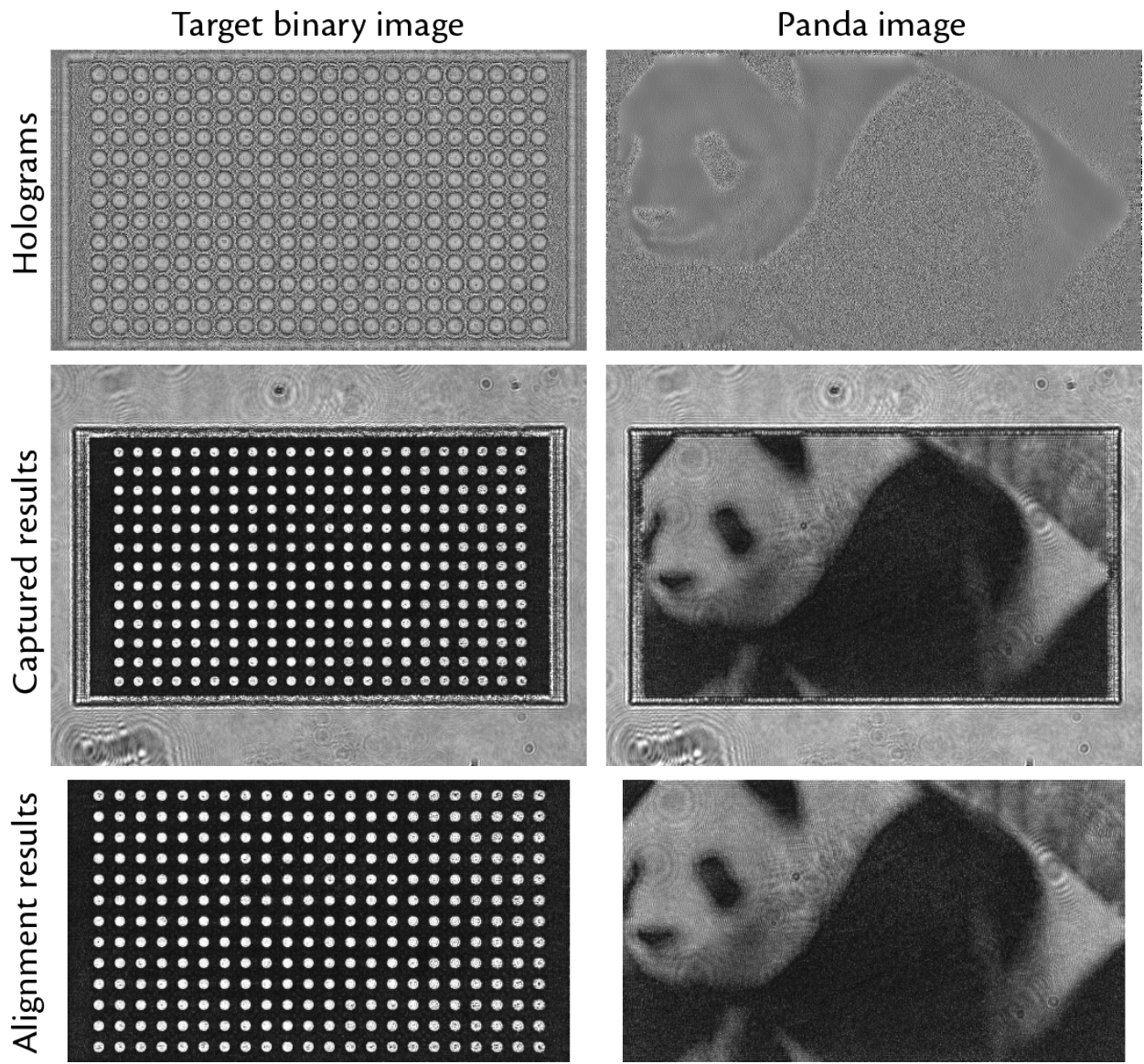

**Figure S11.** Illustration of the calibration images. Image Credits: Eirikur Agustsson and Radu Timofte, ETH Zurich

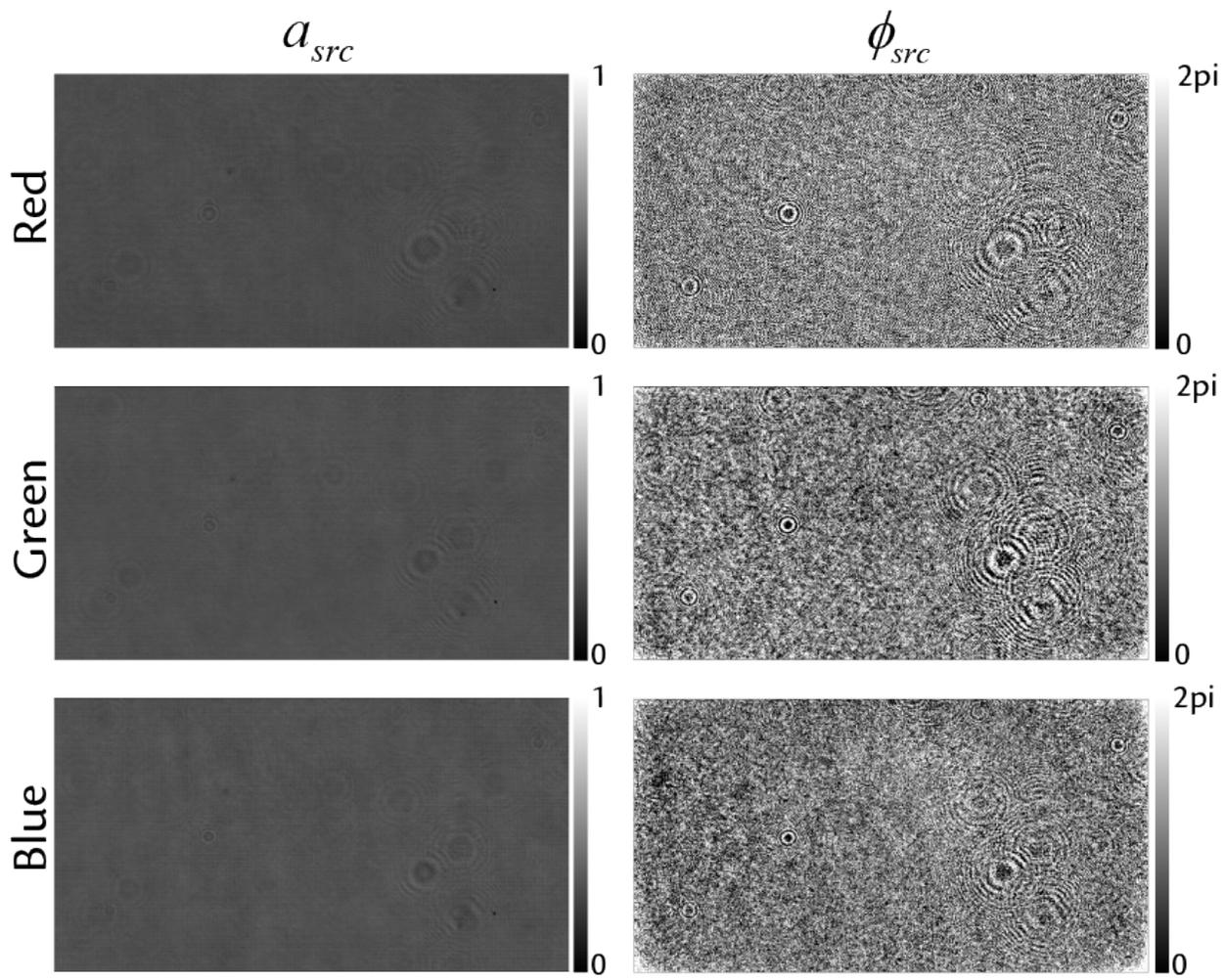

**Figure S12.** Visualization of the physical parameters in the CITL-calibrated model.

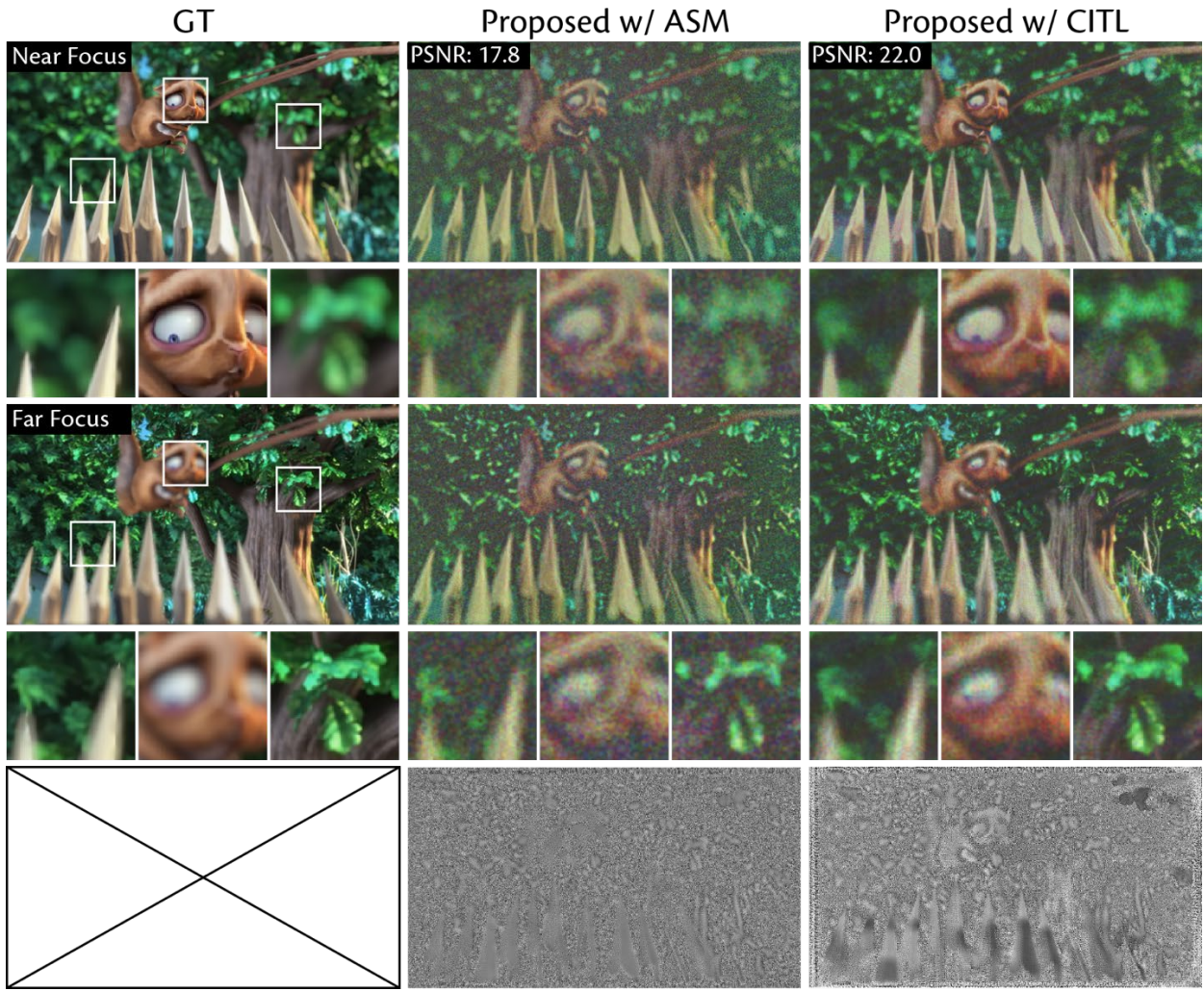

**Figure S13.** Experimental results of the Proposed w/ ASM and CITL. Images Credits: Big Buck Bunny, Blender Institute.

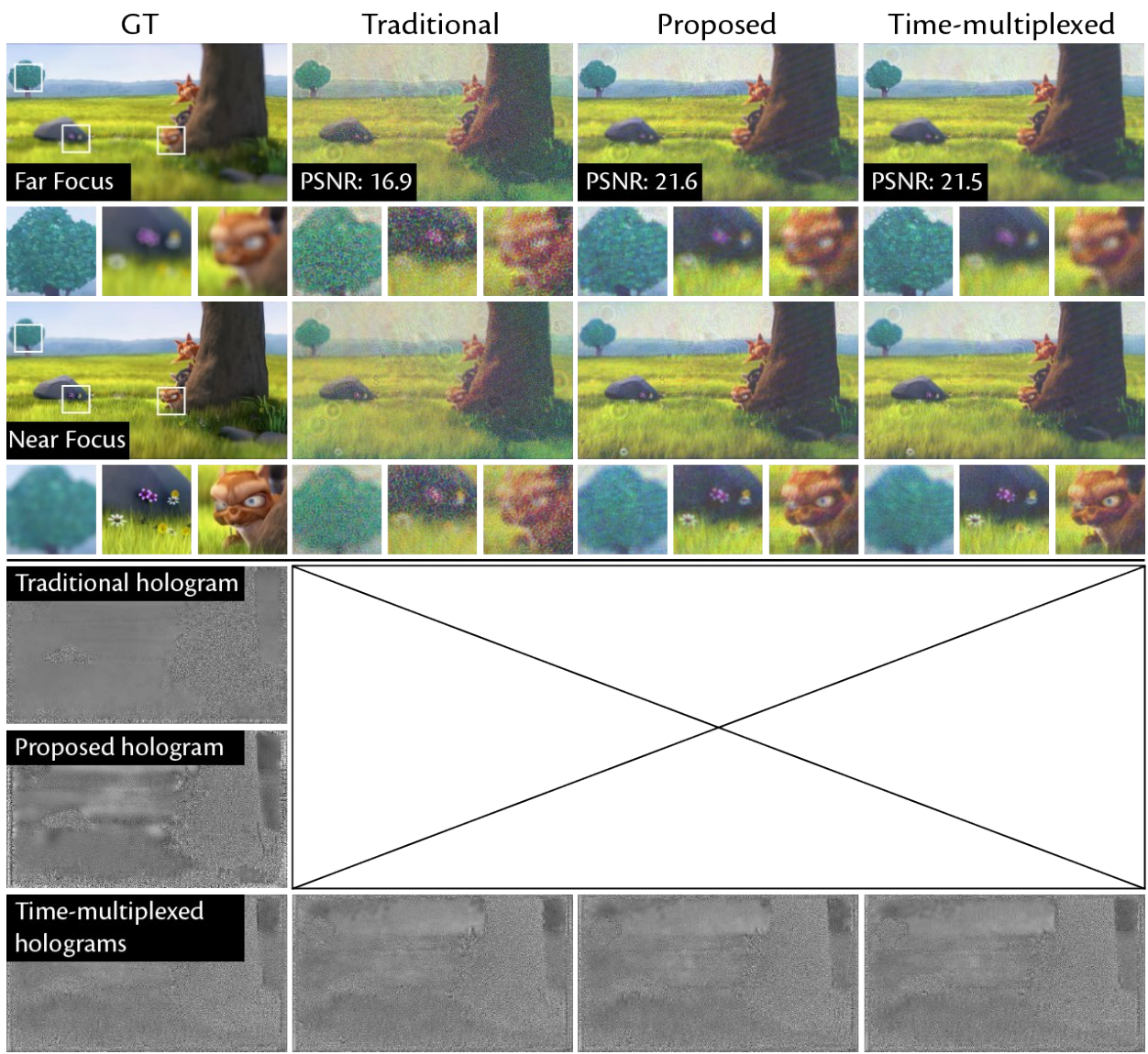

**Figure S14.** Experimental results of images reconstructed with different methods. Image Credits: Big Buck Bunny, Blender Institute.

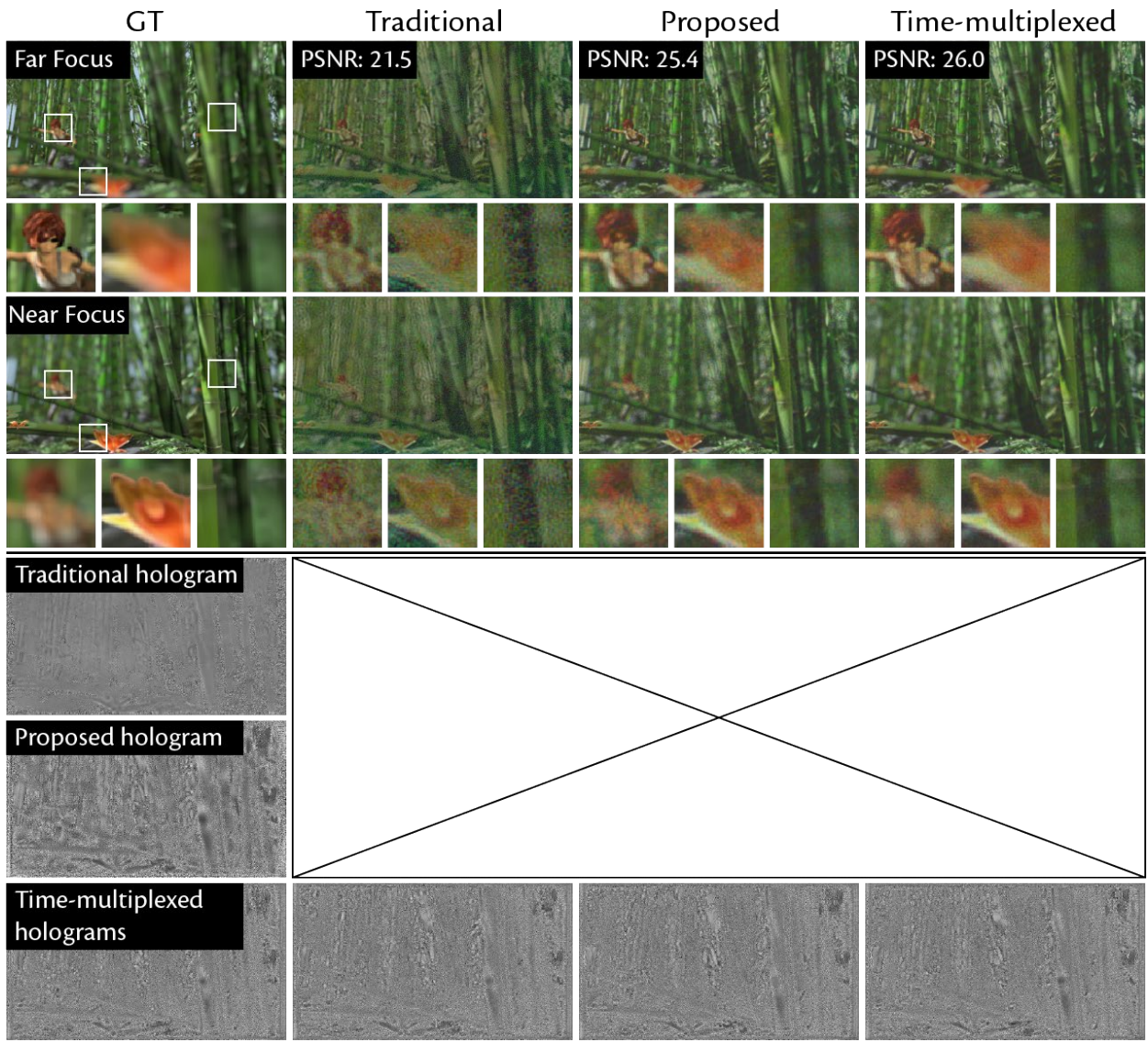

**Figure S15.** Experimental results of images reconstructed with different methods. Image Credits: Daniel J Butler, University of Washington

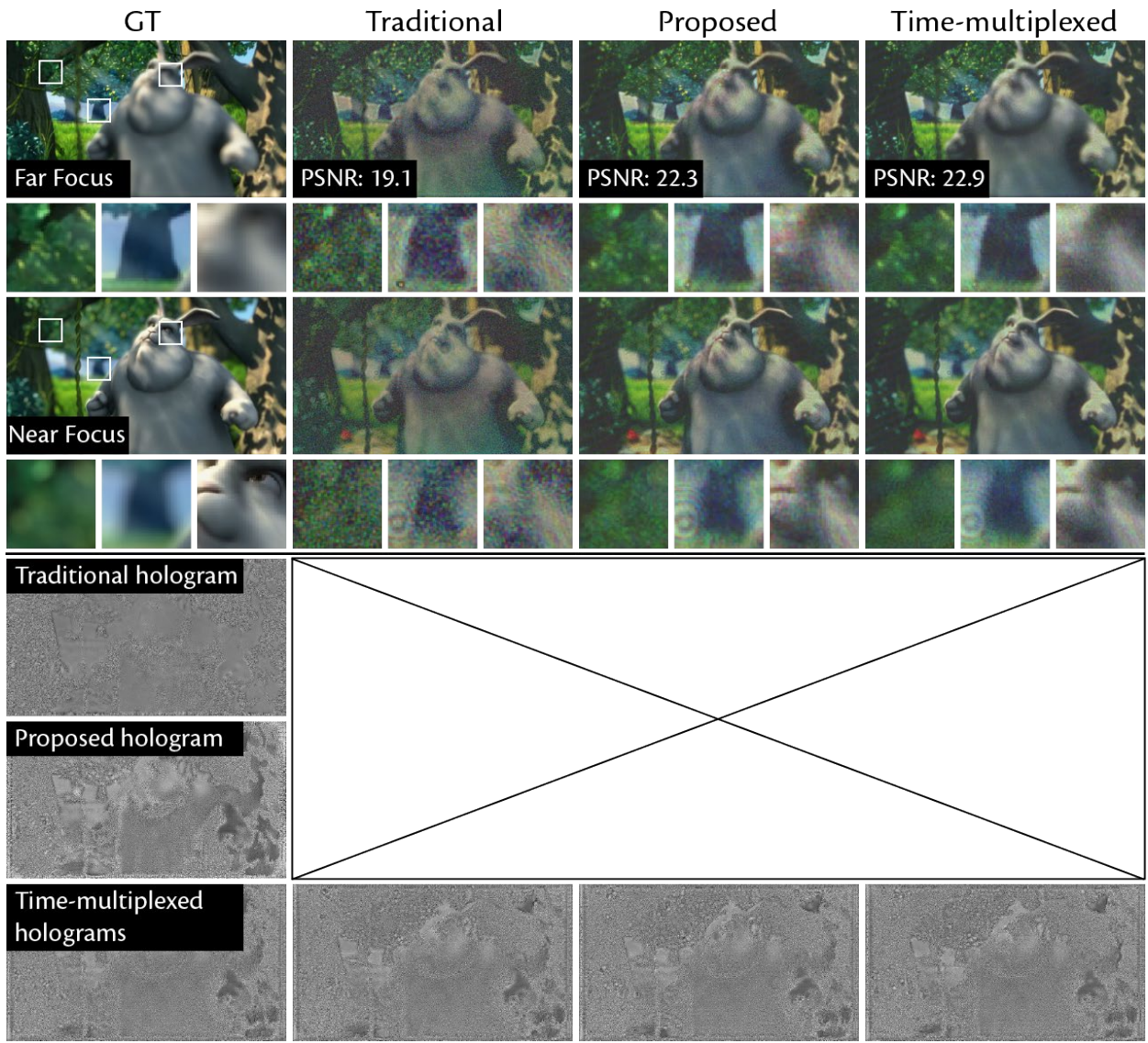

**Figure S16.** Experimental results of images reconstructed with different methods. Image Credits: Big Buck Bunny, Blender Institute.